\documentclass[journal=esthag,manuscript=article,keywords=true]{achemso}
\setkeys{acs}{keywords=true}
\usepackage[version=3]{mhchem} 
\usepackage[numbers]{natbib}
\usepackage[hidelinks]{hyperref}
\usepackage{graphicx} 
\usepackage{tabularx}
\usepackage{array}
\usepackage{xcolor}

\definecolor{darkgreen}{RGB}{0,100,0}

\author{Yang Yan}

\affiliation[OU]
{School of Electrical and Computer Engineering, University of Oklahoma, Norman, OK, 73019 USA}

\author{Zifan Zhou}

\affiliation[PSU]
{Department of Electrical Engineering, Pennsylvania State University, University Park, PA, 16801 USA}

\author{Xuan Wang}

\affiliation[PSU]
{Department of Electrical Engineering, Pennsylvania State University, University Park, PA, 16801 USA}

\author{Erum Hassan}

\affiliation[OU]
{School of Electrical and Computer Engineering, University of Oklahoma, Norman, OK, 73019 USA}

\author{Bilguunzaya Mijiddorj}

\affiliation[OU]
{School of Electrical and Computer Engineering, University of Oklahoma, Norman, OK, 73019 USA}

\author{Jie Cao}

\affiliation[OUCS]
{School of Computer Science, University of Oklahoma, Norman, OK, 73019 USA}

\author{Bin Li}

\affiliation[PSU]
{Department of Electrical Engineering, Pennsylvania State University, University Park, PA, 16801 USA}

\author{Binbin Weng}

\affiliation[OU]
{School of Electrical and Computer Engineering, University of Oklahoma, Norman, OK, 73019 USA}
\email{binbinweng@ou.edu}

\title[An \textsf{achemso} demo]
  {A Locally Deployable Tool-Grounded LLM Multi-agent Framework for Automating Methane Emission Analysis and Reporting}

\abbreviations{IR,NMR,UV}

\begin{document}



\begin{abstract}

Methane field monitoring requires the integration of sampling design, meteorological interpretation, sensor processing, plume analysis, visualization, and reporting, but these steps are often distributed across separate expert-driven workflows. We developed a locally deployable, tool-grounded large language model (LLM) multi-agent framework for our low-cost methane sensing and field-monitoring campaigns. The framework uses LLM agents as workflow coordinators that link field measurements, meteorological data, deterministic sensor-processing routines, Gaussian plume inversion, and report generation, rather than directly estimating methane concentrations or emissions. Extensive field deployments across diverse real-world environments (e.g., wastewater treatment facilities, landfills, and oil and gas sites) demonstrate that our framework can achieve 92.0\% accuracy in workflow routing and parameter extraction, 85.0\% success in emission-rate estimation and plume prediction, and 95.0\% success in generating editable reports under practical operating conditions. Compared with manual and general-purpose LLM-assisted workflows, it reduced workflow time from hours-level to minutes-level, lowered manual coordination and prompt-engineering requirements, and retained traceable plume-based outputs. In addition, most processing can be performed locally, reducing exposure of sensitive facility and field data to cloud services. These results indicate that tool-grounded LLM coordination can reduce the time, labor, usability, and data-security barriers of methane field monitoring.

\end{abstract}

\section{Introduction}
Methane (CH$_4$) is a short-lived greenhouse gas with a strong warming effect and reducing methane emissions is therefore an important part of efforts to limit near-term climate warming
\cite{nisbet2020methane,saunois2025global,jackson2024human}.
Beyond its climate impact, methane leakage also raises safety concerns because the gas is highly flammable and can accumulate to hazardous levels in confined or poorly ventilated spaces
\cite{duncan2015methane}.
Anthropogenic methane emissions arise largely from fossil fuel production and agriculture, with landfills and wastewater treatment facilities also contributing important local sources
\cite{saunois2025global,jackson2024human}.
At these sites, emissions can be intermittent and vary considerably across relatively short distances, which makes them difficult to characterize using sparse or infrequent measurements
\cite{brandt2014methane,alvarez2018assessment,chen2023assessing,schwietzke2017improved,wang2022multiscale}.
More detailed measurements at individual facilities are therefore needed to determine where methane is released and to estimate the associated emission rate
\cite{erland2022recent,chen2023assessing,bell2023performance}.

\begin{figure}[h]
    \centering
    \includegraphics[width=1.0\linewidth]{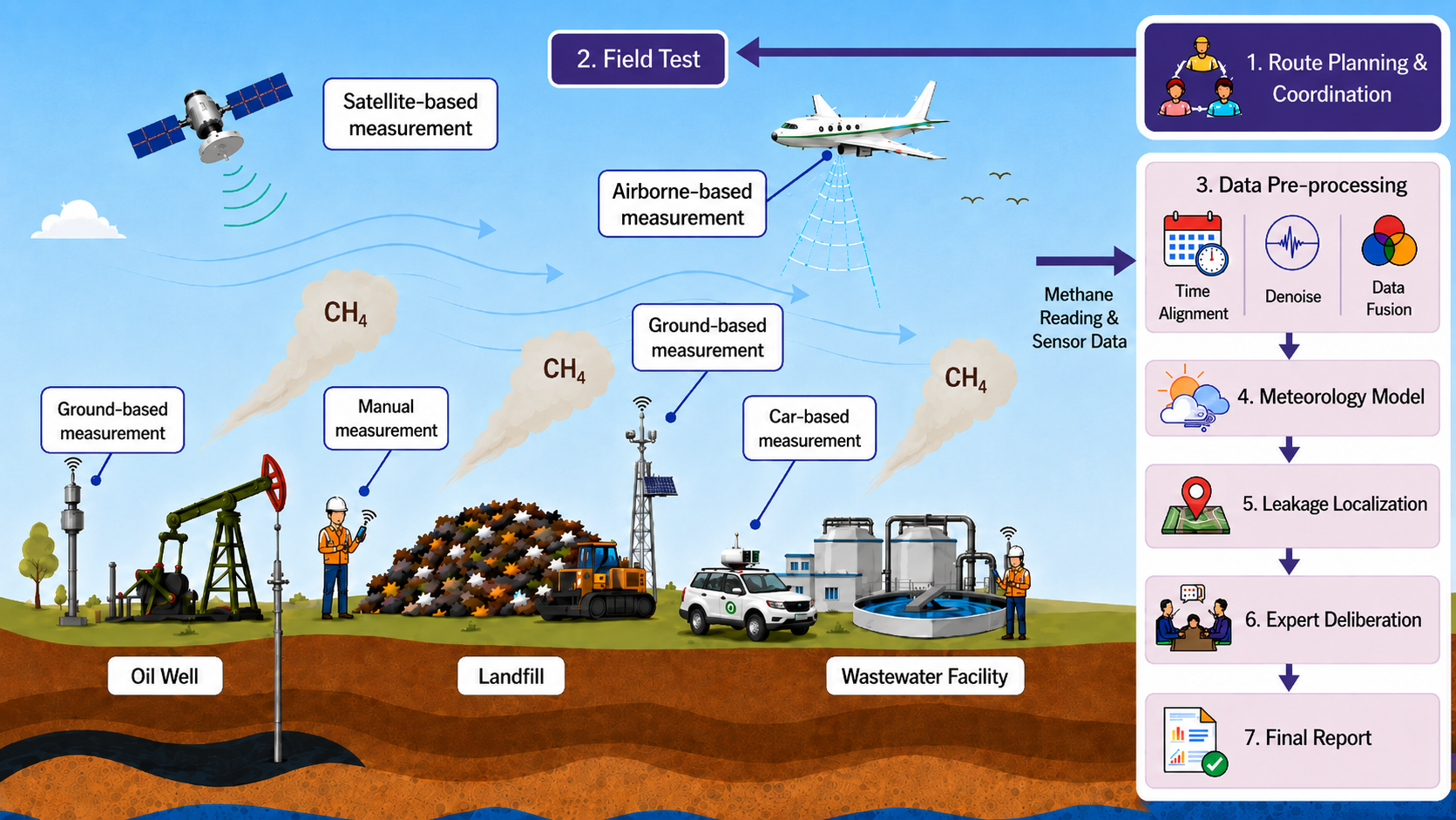}
    \caption{Conventional multi-expert methane monitoring workflows. The figure was generated with AI assistance based on author-provided content and subsequently reviewed and refined by the authors.}
    \label{fig:workflow_comparison}
\end{figure}

As shown in Figure~\ref{fig:workflow_comparison}, current methane monitoring mainly relies on satellite observations, airborne surveys, and ground-based measurements
\cite{jacob2016satellite,erland2022recent,sherwin2022single,thorpe2023attribution}.
Satellite and airborne observations provide broad spatial coverage, whereas ground-based and mobile measurements are needed when the objective is to examine individual sites more closely, locate emission sources, or conduct repeated surveys
\cite{ijzermans2024long,chen2023assessing,ball2025performance,metzger2025framework,coburn2018regional,yang2025assessing,zhou2021mobile}. Vehicle-based methane surveys can further provide rapid spatial coverage for leak detection, source localization, and emission-rate estimation.\cite{weller2018vehicle}
To support these field measurements, our previous studies developed AIMNet, a low-cost nondispersive infrared methane sensing platform that has been used for fixed-site monitoring as well as vehicle-based and portable surveys
\cite{zhou2026aimnet,yan2025machine,yan2025development,hu2026multi,weng2024road}.
However, substantial manual effort is still required before and after field measurements because the measurement plan must be prepared in advance and the collected data still need to be processed afterward. 

In practice, field planning and post-measurement analysis are often handled using separate tools and procedures. For example, locating a suspected methane source first requires the operator to determine where measurements should be collected based on the site and weather conditions. After the survey, the measurement data may need to be organized, aligned with meteorological records, converted to consistent units, and corrected for background concentration before source localization or emission-rate estimation can be performed. Previous studies have developed methods for individual stages of this process, including monitoring design, source attribution, and emission quantification
\cite{metzger2025framework,coburn2018regional,ball2025performance}.
Combining these stages into a complete workflow can still require users to transfer data between tools and maintain consistent analysis settings. This becomes particularly inconvenient when field objectives are given as simple natural-language requests rather than model-ready inputs. A framework that can interpret these requests and coordinate the required tools could therefore reduce the manual effort between field measurement and final analysis.

Large language models (LLMs) have demonstrated strong capabilities in interpreting natural-language instructions
\cite{vaswani2017attention,brown2020language,achiam2023gpt,minaee2024large}.
More recent agentic frameworks extend these capabilities by allowing LLMs to coordinate external tools and specialized workflow components
\cite{wu2024autogen,guo2024large,duan2024exploration,han2024llm}.
Tool-grounded multi-agent systems provide a way to separate these functions by assigning each agent a specific task and a predefined set of tools or data sources.
The LLM can then interpret the request and coordinate the workflow while quantitative analysis is performed by the corresponding processing routines and physics-based models.

Recent studies have begun to apply LLMs and LLM agents to environmental data analysis, decision support, and geospatial applications\cite{zhai2026waterrag,li2026natural,cheng2026leveraging,chen2026leveraging}. However, reliable tool execution and domain-specific quantitative reasoning remain important challenges in such applications. In addition, field sensing requires the coordination of measured data, meteorological information, and physical analysis models rather than the interpretation of existing digital data alone. Integrating these elements into an agentic workflow remains an important step toward practical environmental monitoring.

In this work, we apply this tool-grounded design to methane field monitoring by developing a locally deployable LLM multi-agent framework using the AIMNet platform.
As shown in Figure~\ref{fig:agent_architecture}, the framework begins with a natural-language request and first identifies the type of monitoring task. 
It can then support measurement planning or process collected methane data for further analysis. For source-analysis tasks, deterministic plume-modeling tools are used to reconstruct the methane plume and estimate the likely leak location and emission rate, and the resulting analysis is organized into a standardized field report. 
To evaluate the framework under realistic conditions, we deployed it in wastewater, landfill, and oil and gas environments and assessed its performance across the complete monitoring workflow.
Unlike conventional methane monitoring workflows that require analysts to connect these steps manually, the proposed framework coordinates them within a single local system while retaining established processing routines and physical models for quantitative analysis.
Field evaluation shows that this integrated workflow substantially reduces manual effort and processing time while maintaining reliable quantitative performance.

\section{Overall Multi-agent System Description}

An LLM agent is a software component built around a large language model and assigned a specific role within a workflow. Its behavior is defined by task instructions together with the tools and structured information available to it, allowing the agent to receive a task and return results in a form that can be used by other parts of the system. A multi-agent system extends this structure by coordinating several agents, with each agent handling a different part of the overall task and passing intermediate results to other agents when needed. By separating a complex workflow into well-defined stages, the system allows each agent to focus on a limited responsibility while intermediate outputs can be checked before they are used in subsequent steps
\cite{wu2024autogen,li2023camel,hong2023metagpt,guo2024large}.

\begin{figure}[h]
    \centering
    \includegraphics[width=1.0\linewidth]{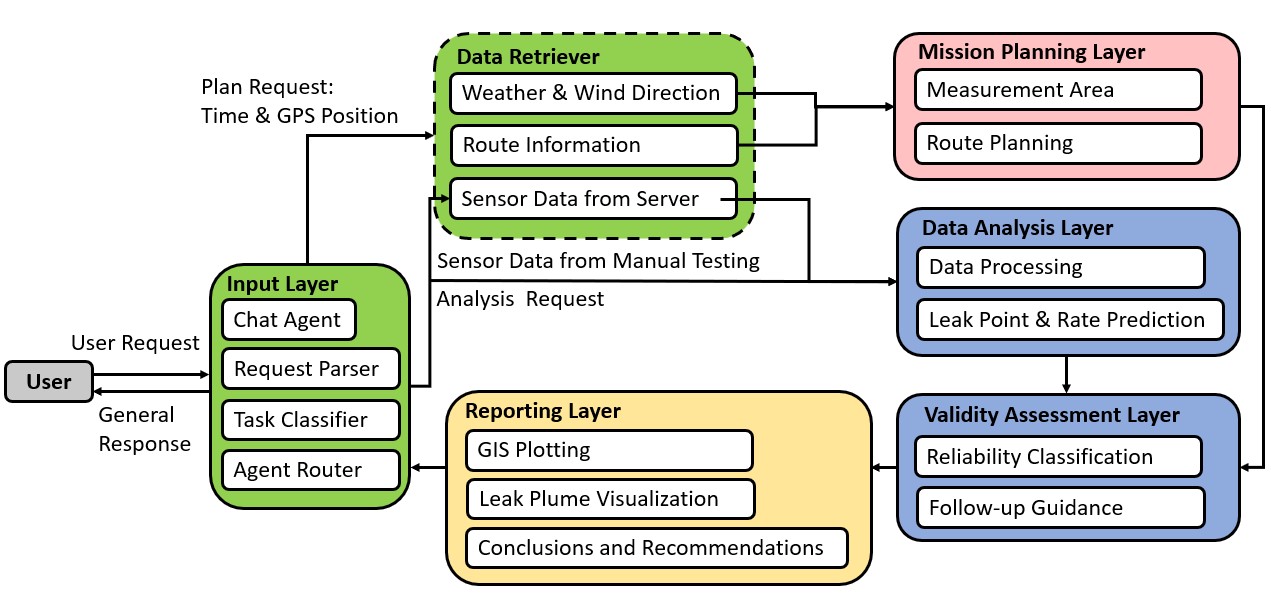}
    \caption{System architecture and workflow of the tool-grounded LLM multi-agent framework for methane monitoring.}
    \label{fig:agent_architecture}
\end{figure}

Figure~\ref{fig:agent_architecture} shows how this multi-agent structure is applied to methane monitoring. 
The framework is organized into four layers that move from user input and task routing through mission planning and methane data analysis to decision support and report generation. 
Within this structure, each agent is assigned a limited role and exchanges structured information with other agents or deterministic tools as the task progresses. 
The LLM agents use the user request to determine the required task and parameters, then coordinate the corresponding workflow and pass the resulting information to the next stage. Quantitative methane analysis is not performed directly by the LLM. Instead, sensor-processing routines and physics-based plume models are used to generate the numerical results. 
This separation allows the LLM to coordinate the overall workflow without replacing the established calculations used for methane analysis, keeping the quantitative results traceable and physically grounded.

Local deployment was adopted because methane monitoring data may contain sensitive facility and location information\cite{huang2025middle}. Reliable Internet access may also be unavailable during field campaigns, making local execution advantageous for field deployment.

In this study, the framework was deployed on a workstation running 64-bit Microsoft Windows 11.
The workstation was equipped with an Intel Core i9-13900KF CPU with 24 cores and 32 logical processors, 64 GB of RAM, an NVIDIA GeForce RTX 4090 GPU with 24 GB of VRAM, and approximately 3 TB of SSD storage.
The local models were served through Ollama \cite{ollamaGithub2026}. Llama~3.1 8B was used as the default model for general interaction and task routing. Qwen 30B was used for more complex text reasoning tasks. Qwen-VL 8B was used for image interpretation.
\cite{grattafiori2024llama,ollamaGithub2026,yang2025qwen3,bai2025qwen3vl}.
User prompts, sensor measurements, and intermediate analytical results therefore remained within the local computing environment. External weather and mapping services were queried only when required. The framework can also support other locally deployed open-weight models, including the Llama and Qwen families
\cite{grattafiori2024llama,yang2025qwen3},
or commercial LLM services when cloud-based inference is preferred.

\subsection{Data Input Layer}

As shown in Fig.~\ref{fig:agent_architecture}, the input layer serves as the entry point between the user and the downstream task-specific agents. Users can describe a monitoring task in natural language or upload a CSV file containing field measurements and their associated location and time information. These inputs can be used for general inquiries, measurement planning, or analysis of previously collected methane data.

The Request Parser and Task Classifier converts the user input into a structured task specification that identifies what the user is asking, where and when the task applies, and what information is needed to complete it. The Agent Router then uses this specification to direct the request to the appropriate downstream agent and analytical tools.

\begin{figure}[htbp]
    \centering
    \includegraphics[width=0.7\linewidth]{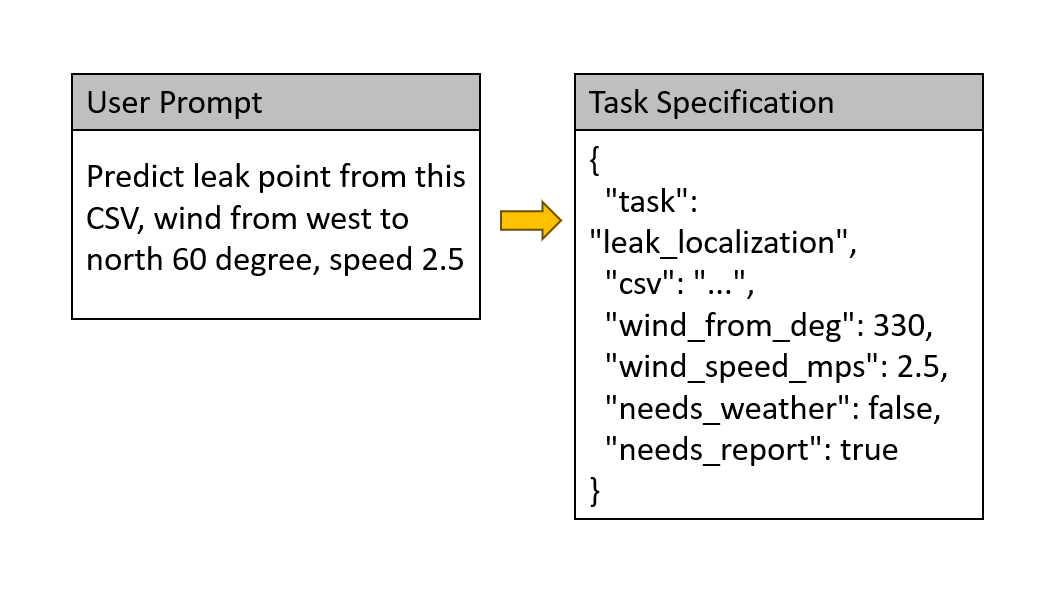}
    \caption{Example conversion of a natural-language request into structured parameters for methane leak localization.}
    \label{fig:prompt_conversion}
\end{figure}

As illustrated in Fig.~\ref{fig:prompt_conversion}, planning and analysis requests are converted into a machine-readable format with standardized fields and units. General questions are routed directly to the selected LLM, whereas task-oriented requests may invoke Google Maps for geographic information, Open-Meteo for meteorological data, and the Message Queuing Telemetry Transport (MQTT)-based AIMNet database for methane measurements
\cite{zhou2026aimnet}.

\subsection{Mission Planning Layer}

The mission planning layer is designed to support efficient methane emission monitoring before field deployment. In conventional field measurements, operators usually carry the sensing device around the target facility and repeatedly search the downwind area to locate potential emission sources. Although this strategy is practical, it can be time consuming because the effective search region is strongly influenced by the target location, site layout, wind direction, wind speed, and short-term meteorological fluctuations\cite{albertson2016mobile}.

To address this limitation, the mission planning agent extracts the planned measurement time, target location, and facility type from the user request. It retrieves the meteorological conditions for the site and uses the current wind speed and direction to estimate the downwind region where methane enhancement is most likely to be detected. The agent also considers forecast wind conditions over the following three hours, which approximates the upper duration of a typical field survey in this study, and adjusts the search region according to the expected wind trend. By accounting for both the current conditions and their expected changes during the survey, the resulting plan identifies the most favorable area for methane measurements and reduces unnecessary searching in the field.

Based on the estimated monitoring region, the path planning agent generates an executable route for field deployment. The agent combines the mission planning output with GPS route information, road accessibility, and site geometry to identify efficient driving or walking paths. It further highlights high-probability detection areas where repeated measurements are recommended. 

\subsection{Data Analysis Layer}

For analysis-oriented tasks, the Data Analysis Agent retrieves methane measurements either from deployed sensing nodes through MQTT or from user-uploaded datasets collected using portable or vehicle-based sensing platforms. The input datasets may contain methane concentrations, GPS coordinates, timestamps, meteorological variables, and source information. The agent then coordinates data preprocessing and invokes the task-specific analytical routine.

For GPS-referenced mobile measurements, the preprocessing routine checks the coordinate and timestamp fields and identifies the methane concentration field before analysis. Measurement units are converted to a consistent basis, and the background methane concentration is estimated from the observed data when no value is provided. Column names may differ among sensing instruments and processing workflows, so the methane field is selected based on the available metadata and expected naming conventions. Distribution-based checks are used when the field remains ambiguous. If the user specifies a methane column, that field is used after validation; otherwise, the most likely field is selected automatically.

The numerical distribution and available metadata are subsequently evaluated to distinguish ppm- and ppb-scale measurements. When the median methane value is below 100, the measurements are interpreted as ppm-scale values unless the column name or metadata indicate ppb-scale concentrations or enhancements. If no background concentration is provided, the background methane level is estimated from the lower portion of the observed concentration distribution.

Meteorological parameters are obtained using a priority-based procedure. User-provided wind speed, wind direction, and atmospheric stability class are used when available. If these parameters are incomplete but valid GPS and time information are provided, the Weather Agent retrieves wind speed, wind direction, temperature, humidity, and related variables from the selected meteorological source. Atmospheric stability is then assigned using a deterministic rule-based classifier based on wind speed, time of day, and other available weather conditions. If the required meteorological information remains unavailable, predefined fallback parameters are applied and the resulting analysis is assigned lower validity. For known-source plume reconstruction, missing meteorological parameters can alternatively be inferred by fitting candidate plume conditions to the observed methane measurements.

In this study, a steady-state Gaussian plume model was used for source localization, emission-rate estimation, and plume reconstruction. The model was selected because it is widely used and computationally efficient for demonstrating the end-to-end analysis workflow. Other dispersion or inversion methods can also be incorporated into the same analysis layer when needed.

The steady-state Gaussian plume concentration is expressed as

\begin{equation}
C(x,y,z) =
\frac{Q}{2\pi u \sigma_y \sigma_z}
\exp\left(-\frac{y^2}{2\sigma_y^2}\right)
\left[
\exp\left(-\frac{(z-H)^2}{2\sigma_z^2}\right)
+
\exp\left(-\frac{(z+H)^2}{2\sigma_z^2}\right)
\right],
\end{equation}
where $C(x,y,z)$ is the methane concentration enhancement at the receptor location. $Q$ is the emission rate, $u$ is the wind speed, and $H$ is the effective source height. The variables $x$, $y$, and $z$ represent the downwind, crosswind, and vertical distances from the source. The parameters $\sigma_y$ and $\sigma_z$ represent the lateral and vertical dispersion coefficients. They are determined from the downwind distance and the assigned atmospheric stability class. In this study, $C$ represents methane enhancement above the estimated background concentration rather than the absolute methane concentration.

For GPS-referenced mobile measurements, the receptor height is denoted as $z_r$, and the effective source height may be specified by the user or approximated based on the facility configuration. The processed measurements are then used to jointly estimate the source location and emission rate. For a candidate source location $\mathbf{s}$, the predicted methane concentration at the $i$-th measurement point is represented as

\begin{equation}
\hat{c}_i(\mathbf{s},Q) = b + Qk_i(\mathbf{s}),
\end{equation}

where $b$ is the estimated background methane concentration, $Q$ is the methane emission rate, and $k_i(\mathbf{s})$ is the unit-emission plume response at the $i$-th receptor for the candidate source location $\mathbf{s}$. The plume response depends on the relative downwind and crosswind positions of the receptor, wind speed, wind direction, atmospheric stability class, receptor height, and effective source height. The term $Qk_i(\mathbf{s})$ therefore represents the modeled methane enhancement above the background concentration.

For each candidate source location, the optimal non-negative emission rate is estimated by weighted least squares:

\begin{equation}
Q^{*}(\mathbf{s}) =
\arg\min_{Q \geq 0}
\sum_{i=1}^{N}
w_i
\left[
c_i-b-Qk_i(\mathbf{s})
\right]^2,
\end{equation}

where $c_i$ is the observed methane concentration at the $i$-th measurement point, $N$ is the number of measurements, and $w_i$ is a data-dependent weight. Methane-enhanced observations are assigned greater weights to emphasize plume-affected measurements, while background-level observations are retained as negative evidence. This treatment penalizes candidate sources that would predict strong methane enhancements at locations where no corresponding enhancement was observed.

Because the Gaussian plume response is linear with respect to $Q$, the optimal emission rate for a given candidate source can be obtained as

\begin{equation}
Q^{*}(\mathbf{s}) =
\max\left[
0,
\frac{
\sum_{i=1}^{N}
w_i k_i(\mathbf{s})(c_i-b)
}{
\sum_{i=1}^{N}
w_i k_i^2(\mathbf{s})
}
\right].
\end{equation}

The fit associated with each candidate source is evaluated using the weighted root-mean-square error,

\begin{equation}
\mathrm{WRMSE}(\mathbf{s}) =
\sqrt{
\frac{
\sum_{i=1}^{N}
w_i
\left[
c_i-b-Q^{*}(\mathbf{s})k_i(\mathbf{s})
\right]^2
}{
\sum_{i=1}^{N} w_i
}
},
\end{equation}

and the final source location is selected as

\begin{equation}
\mathbf{s}^{*} =
\arg\min_{\mathbf{s}}
\mathrm{WRMSE}(\mathbf{s}).
\end{equation}

Source localization is conducted using a two-stage inversion procedure. First, a coarse spatial grid search is performed around the mobile measurement path to identify candidate source regions. At each grid point, the unit-emission plume response is calculated, the corresponding optimal emission rate is estimated, and the weighted residual error is recorded. Second, the most promising candidate regions are refined using numerical optimization, allowing the source location and emission rate to vary continuously beyond the initial grid resolution. This procedure reduces computational requirements while retaining continuous source-location estimates.

To characterize localization uncertainty, the spatial error surface is converted into a relative source-likelihood distribution. Candidate locations with lower weighted residual errors are assigned greater relative likelihood, and the normalized distribution is used to determine the reported source region and confidence score. These quantities describe model-based spatial uncertainty under the assumptions of the Gaussian plume formulation. The resulting output includes the predicted source coordinate, model-based emission-rate estimate, estimated background concentration, weighted error metrics, source-likelihood information, and plume visualization.

For surveys containing multiple spatially separated methane enhancements, the analysis routine first calculates the methane enhancement at each GPS location relative to the estimated background concentration. Elevated observations are identified using an adaptive threshold based on the background level and the observed concentration distribution and are then clustered in projected spatial coordinates.

If multiple clusters are separated by more than a predefined distance and each contains a sufficient number of elevated observations, the dataset is treated as a multiple-source case. Source localization is then performed independently for each cluster. The final output includes an overview of the detected methane-enhancement regions together with cluster-level source estimates, plume visualizations, and diagnostic results. If no robust spatially separated clusters are identified, the standard single-source localization workflow is used.

The same analysis layer also supports plume reconstruction when the source coordinate is known but meteorological information is incomplete. In this case, the source location is fixed, and candidate wind directions, wind speeds, and atmospheric stability classes are evaluated. For each candidate meteorological configuration, the Gaussian plume response is calculated at the observed measurement locations, and the corresponding emission rate is estimated using weighted least squares. The configuration that minimizes the weighted RMSE between the modeled and observed methane concentrations is selected as the best-fitting plume condition.

The inferred wind direction, wind speed, and atmospheric stability class are reported as model-derived fitting parameters rather than direct meteorological measurements. These parameters represent the conditions that best reproduce the observed methane distribution under the Gaussian plume assumptions and do not replace independent weather observations.

\subsection{Decision Support and Report Generation Layer}

After the localization, emission estimation, and plume reconstruction workflows are completed, the final outputs are processed by the decision support and report generation layer. This layer is designed to evaluate the reliability of the analytical results, recommend follow-up sampling when needed, and convert intermediate outputs into structured decision support reports. Instead of generating isolated figures or text summaries, the layer integrates historical user inputs, task specifications, processed sensing data, plume prediction results, validity information, generated figures, and previous interaction context to assemble a complete report according to the user's request.

A key component of this layer is the Validity Agent, which evaluates whether the localization or plume reconstruction result is sufficiently constrained by the available data. The validity score is computed from several diagnostic factors, including the RMSE relative to methane enhancement, GPS spatial spread, number of elevated methane points, distance from the predicted source to the sampled route, emission rate magnitude, and source likelihood confidence. The output is a high, medium, or low validity label with explanatory flags. Low validity is assigned when the model residual is high, GPS coverage is too compact, too few elevated points are available, or the predicted source lies far from the sampled route. This validity assessment does not recompute the leak location; instead, it provides an interpretable reliability check for the existing analytical result.

Based on the validity assessment, the framework provides follow-up sampling recommendations to support adaptive field decision making. When the validity is high, the system may recommend an optional confirmation transect to verify the predicted plume structure. When the validity is medium or low, additional sampling is recommended before a final field decision is made. The default recommendation is a cross plume loop around the candidate source, including an upwind background point, two cross plume edge points, and a downwind centerline point. For multi-source cases, the system recommends local verification routes for each detected cluster rather than one global route. In this way, the recommendation module supports practical field planning without claiming fully autonomous route optimization.

The Report Generator Agent then converts the structured results and generated figures into an AIMNet Methane Report. The report can include Geographic Information System (GIS)-based monitoring maps, methane plume visualizations, estimated source locations, emission rate predictions, uncertainty information, validity labels, follow-up sampling suggestions, conclusions, and recommended field actions. Recent results are deduplicated by data source and location so that repeated runs do not dominate the report. Multi-source results are summarized as a single multi-source case, while cluster level outputs are included as supporting evidence when needed. Each figure is paired with a numerical summary and a short visual interpretation.

The Report Generator Agent also supports context-aware figure retrieval and layout refinement. Previously generated plume maps, GIS plots, route planning figures, and analysis tables can be recalled and incorporated into the final report. If the user is not satisfied with the font size, label placement, figure arrangement, caption style, or other formatting details, the feedback can be returned to the agent for iterative revision. User preferred report formats, such as section order, figure style, and conclusion structure, can also be saved and reused in subsequent monitoring tasks to ensure consistent formatting and reduce repetitive manual editing.

When a vision language model is available, it is used only to describe visual context in the generated figures, such as plume footprint, nearby roads, buildings, vegetation, or open land. Numerical methane conclusions, including source location, emission rate, uncertainty, and validity level, are derived from the structured analysis results rather than from the vision model. This design prevents visual interpretation from overriding quantitative plume inversion results while still allowing the report to provide useful spatial context.

In addition to visualization, the agent summarizes the generated data and interprets the analytical results to produce an overall conclusion and recommendation section. This text component remains adjustable by the user, allowing the report to combine automated generation with user directed refinement.

\section{Applications and Discussion}
To evaluate the practical performance of the proposed framework, the developed functions were tested in real-world methane monitoring applications. The application section examines how the system supports key tasks in practice, including planning, data retrieval, methane data analysis, leak localization, emission rate estimation, and report generation. Through field-oriented use cases, the study assesses whether these integrated functions can effectively assist users under realistic operational conditions.

\begin{figure}[H]
    \centering
    \includegraphics[width=1\linewidth]{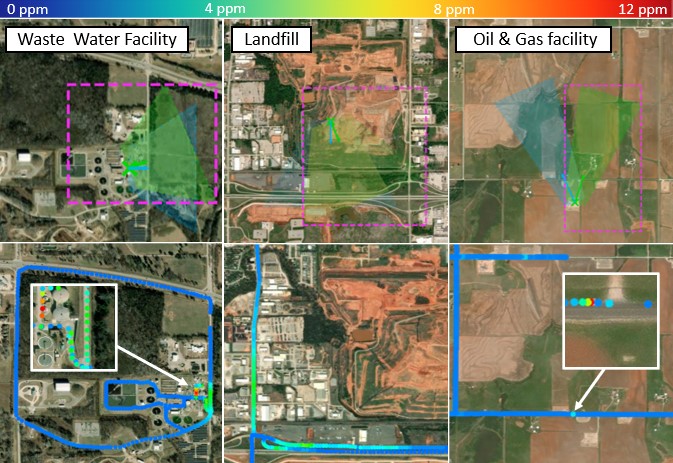}
    \caption{Route-planning outputs generated by the AIMNet agent for three methane monitoring scenarios}
    \label{fgr:route planning agent_1}
\end{figure}

Figure~\ref{fgr:route planning agent_1} presents representative route planning results generated by the proposed framework for methane monitoring under different site conditions. Three field locations were selected for the route planning experiments: a wastewater treatment facility in Norman, Oklahoma, a landfill in Oklahoma City, Oklahoma, and an oil and gas facility in El Reno, Oklahoma. The experiments were conducted at different times to examine the applicability of the proposed framework under different environmental conditions and site layout characteristics.

The field measurements followed our previously established methane monitoring setup, consisting of a vehicle-based platform equipped with the LI-COR LI-7810, LI-COR LI-7700, and AIMNet sensing device to collect methane concentration data along the routes generated by the proposed framework.\cite{hu2026multi} The purpose of this setup was to evaluate whether the planned routes could guide the monitoring platform to areas where methane emission signals from potential leak points could be captured.

The top row shows three example planning scenarios over satellite basemaps. In each case, the system identifies candidate monitoring regions and generates directional search sectors associated with different methane enhancement levels. The color bar, ranging from 0 to 12 ppm, represents the relative methane concentration level considered during the planning process. The highlighted polygons indicate suggested monitoring coverage regions, while the directional sectors represent candidate downwind search orientations for field deployment. The magenta dashed rectangles denote selected target areas used to constrain local route generation and focus subsequent measurements. The green regions represent the currently estimated potential emission areas, whereas the blue regions indicate future or extended search areas predicted by the planning module.

The bottom row illustrates the corresponding route planning outputs for field execution. The blue lines indicate planned driving or traversal paths, while the colored markers represent measurement points or sampled methane observations collected along the route. These examples demonstrate that the proposed framework can adapt route design to different site geometries, including compact industrial facilities, linear roadside corridors, and open field or undeveloped areas. The enlarged insets further show how the planned routes can be refined for local inspection around suspected source regions.

\begin{figure}[H]
    \centering
    \includegraphics[width=1\linewidth]{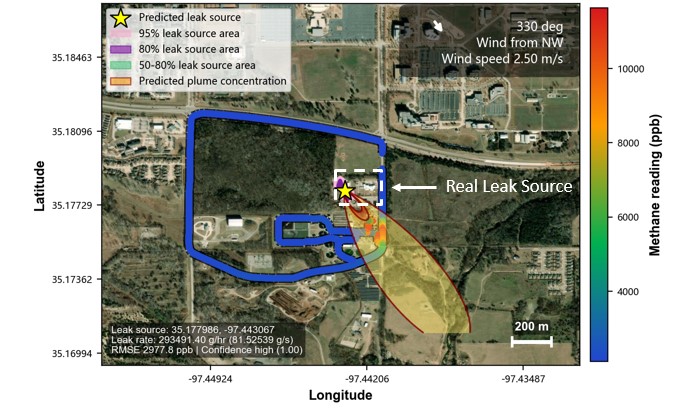}
    \caption{Reverse Prediction of Leak Points from Field Measurement Results at a Wastewater Facility}
    \label{fgr:Leak point prediction}
\end{figure}

Overall, the figure demonstrates that the route planning module integrates site layout, target region selection, wind related search orientation, and methane related spatial cues to generate operationally feasible monitoring paths for field measurements. In the first and third field tests, the major methane enhancement peaks, shown as red points, were located within the overlapping regions of the current and predicted potential emission areas. In the second field test, although part of the observed methane enhancement region was not fully covered by the planned route, the framework still identified the primary emission prone area. This discrepancy may be attributed to local turbulence, temporary wind direction changes, or other short-term meteorological variations during field deployment.

To evaluate the leak source prediction capability of the proposed agent system, a vehicle-based methane survey was conducted around the wastewater facility. Figure~\ref{fgr:Leak point prediction} presents the automatically generated result from an input CSV file containing raw sensor measurements and GPS data. The proposed agent system removes abnormal signals, preprocesses the methane measurements, applies the machine learning model, and generates a GIS-based concentration map. Blue points indicate methane readings near the processed baseline, whereas measurements exceeding the background concentration by at least 1 ppm are selected for reverse prediction of the potential leak area. The resulting 80\% confidence region successfully covered the actual emission points, which were later confirmed through follow-up field detection.

\begin{figure}[H]
    \centering
    \includegraphics[width=1\linewidth]{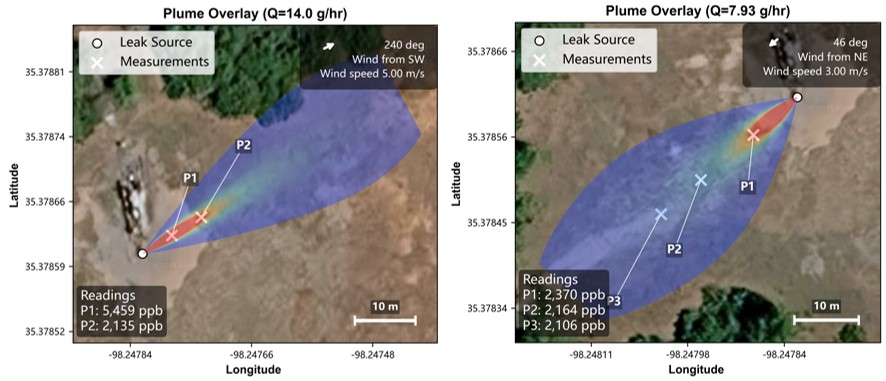}
    \caption{Gaussian plume reconstruction under different meteorological conditions using manual and automated input workflows}
    \label{fgr:Gaussian Plume}
\end{figure}

To validate the emission data analysis and prediction function, the proposed framework was tested using data collected from a real leak monitoring mission. During a field task conducted in collaboration with the Caddo Nation in Oklahoma, methane measurements were manually collected using the LI-COR LI-7810 to investigate abandoned wells with potentially high leakage risk. The selected test location had already been identified as a high methane emission point based on prior field observations, making it a suitable case for evaluating whether the generated plume report was consistent with real field measurements.

Figure~\ref{fgr:Gaussian Plume} presents the quick report generated by the agent. The left panel shows the workflow in which sensor parameters and weather information were manually entered by the user, whereas the right panel shows the automated workflow in which the system extracts sensor readings from the CSV file based on the timestamps and the provided GPS coordinates. In the automated workflow, the system also retrieves the corresponding weather conditions and estimates the atmospheric stability class for plume prediction.

In addition to the input measurement location, multiple manual measurements were collected at other positions within and around the predicted Gaussian plume region for comparison. The observed methane concentration patterns from these manual field measurements were consistent with the plume distribution generated by the agent, indicating that the predicted plume range corresponded well with the area where methane enhancements were detected in the field. Both workflows organize the sensor readings, wind direction, wind speed, sampling labels, measurement positions, and predicted plume distribution into a standardized report format.

To evaluate the repeatability and accuracy of the overall AIMNet agent system, multiple tests were conducted across different facilities under three methane monitoring scenarios, including wastewater facilities, landfills, and two oil and gas facilities. These tests were performed under different time and weather conditions to evaluate the robustness of the system. The evaluation was divided into several workflow level and output quality tasks to assess whether the agent could correctly route user requests, interpret input data, select appropriate environmental parameters, perform methane analysis, and generate usable reports.

As shown in Table~\ref{tab:agent_decision_accuracy}, the proposed agent demonstrated reliable decision making capability across different workflow and data interpretation tasks. For workflow routing and parameter extraction, the agent achieved a success rate of 92.0\% over 50 prompts, indicating that most user instructions were correctly mapped to the corresponding workflow and that the required parameters were successfully extracted. The remaining errors mainly occurred when a CSV file was uploaded without sufficient user instructions. In such cases, the agent occasionally selected an inappropriate workflow and generated an incorrect graph type.

The methane column and unit selection task achieved a success rate of 90.0\%, showing that the agent was generally able to identify the correct methane concentration column and handle ppm/ppb unit conversion. The failed case occurred when the uploaded dataset contained a short data sequence, unusually high concentration values, and missing unit metadata. Under this condition, the agent misinterpreted ppm level values as ppb level measurements.

\begin{table*}[h]
\centering
\caption{Agent decision accuracy across workflow and data interpretation tasks.}
\label{tab:agent_decision_accuracy}
\renewcommand{\arraystretch}{1.18}
\setlength{\tabcolsep}{4pt}
\footnotesize

\begin{tabularx}{\textwidth}{
>{\raggedright\arraybackslash}p{0.22\textwidth}
>{\centering\arraybackslash}p{0.09\textwidth}
>{\raggedright\arraybackslash}p{0.25\textwidth}
>{\centering\arraybackslash}p{0.10\textwidth}
>{\raggedright\arraybackslash}X}
\hline
\textbf{Agent Decision Task} &
\textbf{Test Cases} &
\textbf{Success Criterion} &
\textbf{Success Rate} &
\textbf{Failure Description} \\
\hline

Workflow routing and parameter extraction &
50 prompts &
Correct workflow selected and all required parameters extracted &
92.0\% &
Errors mainly occurred when prompts contained incomplete field descriptions for graphing. \\

Methane column and unit selection &
10 CSV files &
Correct methane column selected and ppm/ppb unit handled correctly &
90.0\% &
One case failed due to unclear column naming and missing unit for high methane concentration \\

Weather source decision &
10 cases &
Correctly selected user provided wind, weather API, inferred wind, or fallback &
100\% &
None \\

Single source vs multi-source decision &
8 CSV files &
Correctly classified the survey as single source or multi-source &
87.5\% &
Errors were mainly caused by overlapping methane enhancement regions from nearby potential sources. \\

Validity decision &
10 cases &
Validity label matched the actual error category &
80.0\% &
Misclassification occurred when multiple error types appeared simultaneously in the same input case. \\

\hline
\end{tabularx}
\end{table*}

The weather source decision task reached 100\%, suggesting that the agent correctly selected among user provided wind information, weather API retrieval, inferred wind settings, and fallback configuration. The single source versus multi-source decision task achieved 87.5\%. The remaining errors were mainly caused by overlapping methane enhancement regions from nearby potential sources, which made it difficult to distinguish whether the observed methane plume was produced by one source or multiple sources.

The validity decision task achieved a success rate of 80.0\%, which was the lowest score among the workflow level decision tasks. This lower performance was mainly due to cases in which multiple error types appeared simultaneously in the same input file or user request. Under these conditions, the agent occasionally assigned the wrong validity label because the input contained more than one possible failure category.

\begin{table*}[h]
\centering
\caption{Methane analysis performance of AIMNet after workflow execution.}
\label{tab:methane_analysis_performance}
\renewcommand{\arraystretch}{1.18}
\setlength{\tabcolsep}{4pt}
\footnotesize

\begin{tabularx}{\textwidth}{
>{\raggedright\arraybackslash}p{0.22\textwidth}
>{\centering\arraybackslash}p{0.09\textwidth}
>{\raggedright\arraybackslash}p{0.25\textwidth}
>{\centering\arraybackslash}p{0.10\textwidth}
>{\raggedright\arraybackslash}X}
\hline
\textbf{Agent Decision Task} &
\textbf{Test Cases} &
\textbf{Success Criterion} &
\textbf{Success Rate} &
\textbf{Failure Description} \\
\hline
Missing wind plume reconstruction &
30 cases &
Correctly used the weather API when time information was available, or applied the default wind setting when detailed time information was missing.&
100\% &
None \\
Potential measurement area prediction &
10 cases &
Correctly cover the potential methane detection area  &
90\% &
Only occurs when wind and environmental conditions are unstable \\

Leak source localization &
15 cases &
Correctly localize the Highest reading point, and Predicted source error $<5$ m & 

73.3\% & Errors mainly occur under unstable or weak wind conditions, which reduce the reliability of plume-based source localization.\\

Emission Rate and Emission plume prediction &
20 cases &
Leak rate estimates and plume visualizations were physically meaningful and consistent with manual calculation. & 

85\% & Incorrect stability class was selected for calculation-based on the weather conditions.\\

Report Generation &
20 cases &
Generated an editable report with meaningful content and structure aligned with the user's request & 

95\% & Failures mainly occurred when the revised report contained corrupted text or misplaced elements\\

\hline
\end{tabularx}
\end{table*}

Table~\ref{tab:methane_analysis_performance} further evaluates the methane analysis performance after workflow execution. The Supporting Information includes representative examples that illustrate the expected output and success criterion for each task. The missing-wind plume reconstruction task achieved 100\%, demonstrating that the agent could correctly use the weather API when time information was available or apply a default wind setting when detailed timing information was missing.

Potential measurement area prediction achieved 90\%, indicating that the agent could generally identify the likely methane detection area. The incorrect predictions mainly occurred under unstable environmental conditions, such as rainy weather or high temperature sunny noon periods, where wind direction and atmospheric conditions were less stable.

Leak source localization achieved 73.3\%, which was the lowest score among the analysis tasks. This result is reasonable because source localization is highly sensitive to wind stability, weak wind conditions, environmental uncertainty, and the spatial coverage of the measurement points. In particular, unstable or low wind conditions can reduce plume consistency, and insufficient test points may further limit the reliability of source localization.

The emission rate and plume prediction task achieved 85\%, showing that the generated leak rate estimates and plume visualizations were generally physically meaningful and consistent with manual calculation. The remaining errors mainly occurred when the agent selected an incorrect atmospheric stability class, which resulted in less meaningful plume calculation results in a small number of cases.

Report generation achieved a success rate of 95\%, indicating that the agent could produce editable reports aligned with the user's request in most cases. The remaining cases produced report files with localized formatting or readability issues, such as corrupted text segments, partially unreadable content, or misplaced elements during revision. Representative examples are provided in the Supporting Information.

\begin{figure}[H]
    \centering
    \includegraphics[width=1\linewidth]{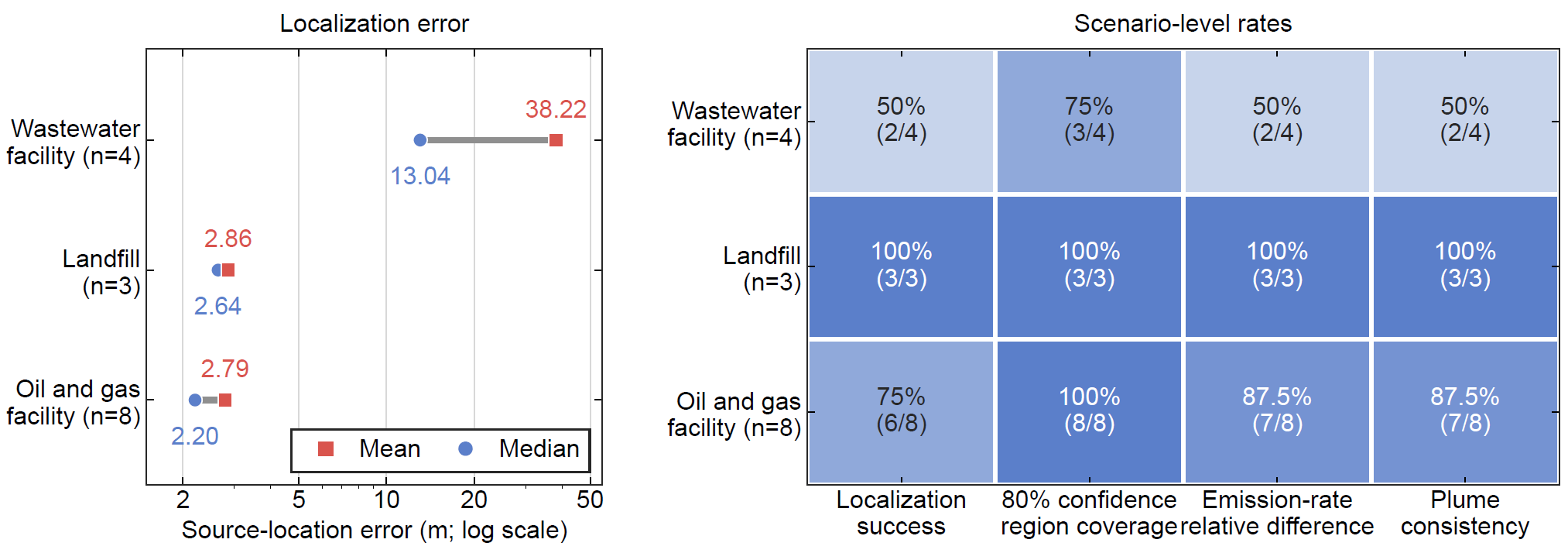}
    \caption{Quantitative validation of leak source localization, emission rate estimation, and plume reconstruction across field scenarios.}
    \label{fgr:leaksource validation}
\end{figure}

Figure ~\ref{fgr:leaksource validation} provides a detailed evaluation of the core methane analysis functions, including source localization, emission rate estimation, and plume prediction. This imbalance mainly reflects differences in monitoring priorities, facility availability, and field campaign logistics. Oil and gas facilities were sampled more frequently because they are a major target for methane detection and are often more numerous and spatially clustered, allowing multiple cases to be collected during one field campaign. In comparison, accessible landfill and wastewater facilities were more limited and subject to site permission, scheduling, and suitable meteorological conditions. Therefore, the landfill and wastewater results should be interpreted as representative case studies rather than comprehensive facility-type evaluations. The results show that the landfill, the oil, and the gas facility cases achieved better overall performance. In the oil and gas facility cases, only one predicted source was outside the expected prediction range, while the mean and median source errors remained relatively low and within an acceptable range for field detection.

In contrast, the wastewater facility cases showed lower localization performance, with a success rate of only 50\%. Two cases produced much larger source errors, which increased both the mean and median source error values. This reduced performance was mainly caused by the complex built environment around the wastewater facility, where nearby buildings, facility structures, and other obstacles can disturb local wind fields and affect methane plume transport.
\cite{yuvaraj2026high} As a result, the Gaussian plume-based localization method is more reliable in open or less obstructed environments, such as landfill and some oil and gas facility scenarios, but it can become less accurate in areas with multiple buildings, trees, or other physical obstructions.

These results indicate that the proposed agent is effective for predicting methane emission sources when the plume transport is not strongly blocked or distorted by surrounding structures. However, in complex environments with dense buildings or vegetation, human review and additional field measurements are still needed to verify the predicted source location. Similar challenges may also occur in future oil and gas field applications if the monitoring area contains trees, equipment clusters, or other obstacles. This limitation could be addressed by integrating more advanced gas dispersion models, such as Gaussian puff models, Lagrangian particle dispersion models, or CFD assisted wind field models, to improve plume prediction and source localization under complex field conditions.

\begin{figure}[H]
    \centering
    \includegraphics[width=1\linewidth]{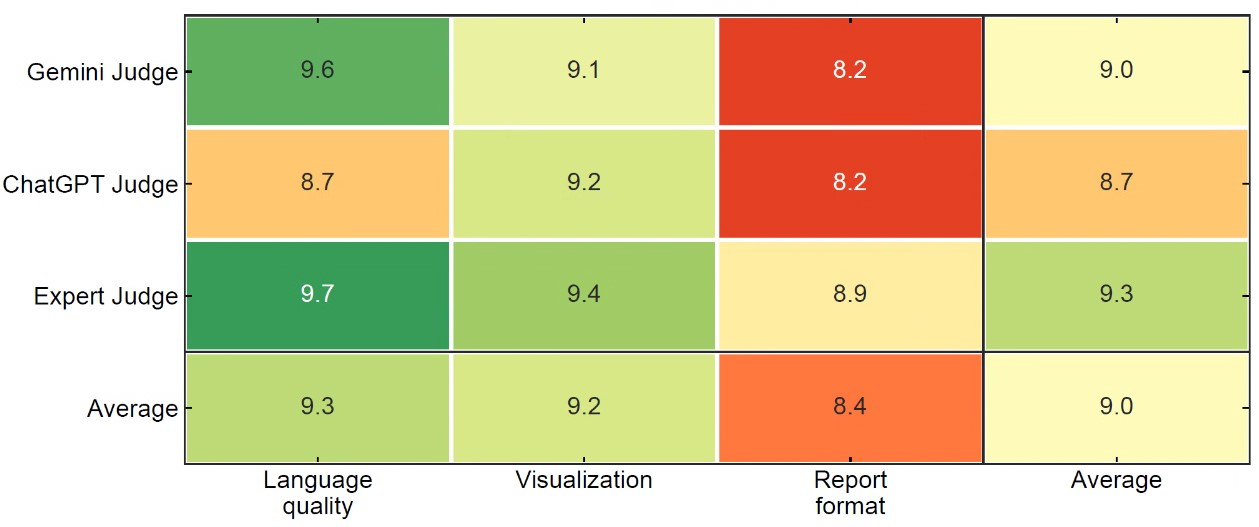}
    \caption{External LLM-based and expert evaluation of the locally deployed AIMNet agent across different report generation tasks}
    \label{fgr:llmscore}
\end{figure}

The external evaluation results in Figure~\ref{fgr:llmscore} show that the generated reports received consistently high scores from Gemini Judge, ChatGPT 5.5 Judge, and an expert judge. The average scores for language quality, visualization, and report format were 9.3, 9.2, and 8.4, respectively, resulting in an overall average score of 9.0. Among the three criteria, language quality and visualization received the highest scores, suggesting that the agent could generate clear technical descriptions and meaningful figures for methane analysis reports. The report format score was slightly lower, mainly because report layout and formatting are more sensitive to figure placement, table alignment, and revision consistency. The expert judge provided the highest overall score of 9.3, which further supports the usability of the generated reports for practical methane monitoring analysis.

\begin{figure}[H]
    \centering
    \includegraphics[width=1\linewidth]{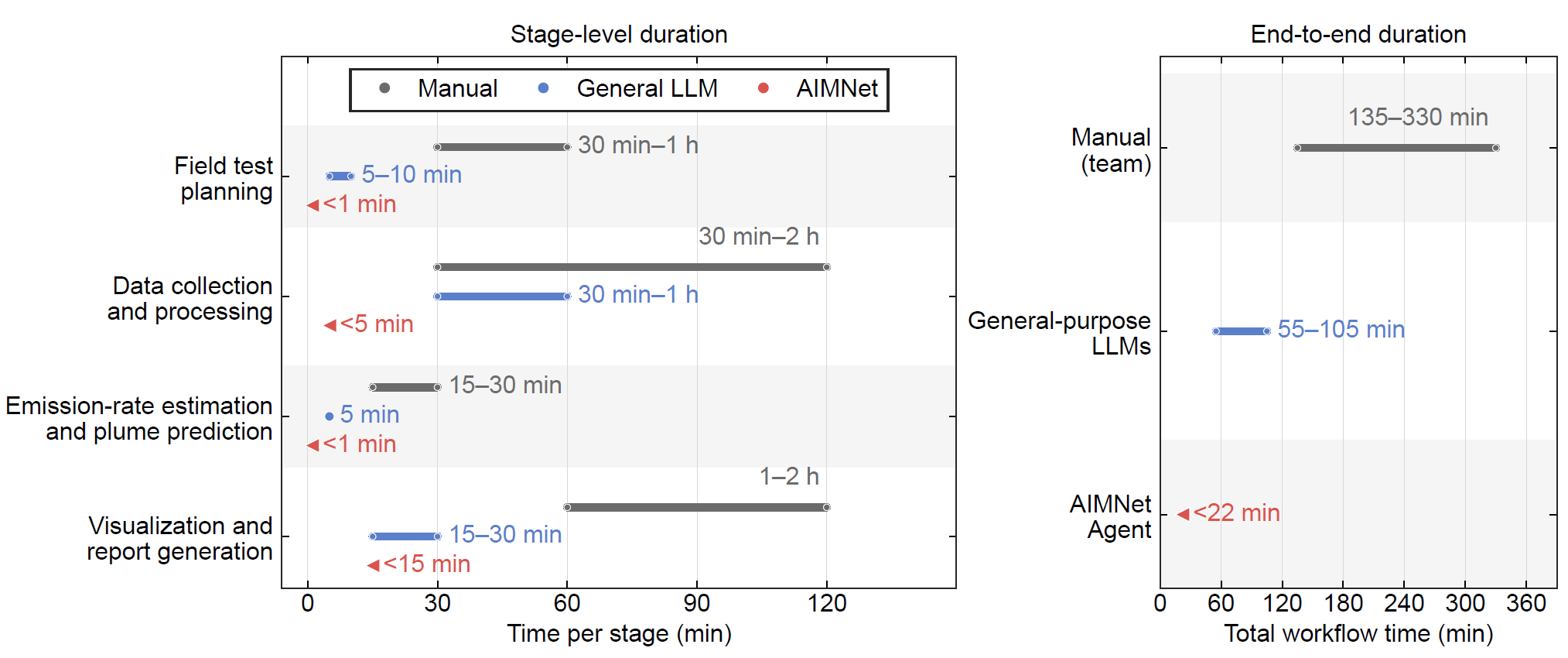}
    \caption{Estimated time comparison among the manual workflow, general purpose LLMs (ChatGPT), and the AIMNet agent}
    \label{fgr:timeComparision}
\end{figure}

Figure~\ref{fgr:timeComparision} compares the estimated time required by the manual workflow, general purpose LLMs (ChatGPT), and the AIMNet agent. In the manual workflow, the task was assumed to be completed by a two to three person expert team through discussion, manual data interpretation, calculation, visualization, and report preparation. For the general purpose LLM workflow, an expert user was still required to provide domain knowledge, interpret the model outputs, and manually refine the analysis results. In contrast, the AIMNet agent was operated by a student user with limited prior experience in methane emission analysis and visualization, demonstrating the potential of the agent to reduce the required level of domain expertise.

For field test planning, the manual workflow required approximately 30 minutes, while general purpose LLMs required 5 to 10 minutes and the AIMNet agent completed the task in less than 1 minute. For data collection and processing, the manual workflow required 30 minutes to 2 hours, compared with 30 minutes to 1 hour for general purpose LLMs and less than 5 minutes for the AIMNet agent. Similar improvements were observed in emission rate estimation, plume prediction, visualization, and report generation. In particular, visualization and report generation were reduced from 1 to 2 hours in the manual workflow to less than 15 minutes using the AIMNet agent.

These results indicate that the proposed agent substantially reduces the time required for methane monitoring workflows while maintaining meaningful analytical and reporting quality. More importantly, the AIMNet agent reduces the dependence on expert level manual operation by integrating workflow routing, data processing, plume-based calculation, visualization, and report generation into a single automated framework.

\section{Conclusion}

The central finding of this study is that LLMs can provide practical value in methane monitoring when they are used to coordinate, rather than replace, deterministic environmental analysis. By translating natural-language requests into structured tasks and connecting sensor processing, meteorological data, plume inversion, visualization, and reporting tools, the proposed framework reduced the manual effort required to move from field measurements to interpretable results. The field evaluations further showed that this tool-grounded design can maintain traceable quantitative outputs while making complex methane analysis more accessible to users with limited prior experience. The main contribution is therefore not the use of an LLM to calculate methane emissions, but the development of a constrained coordination layer that connects otherwise fragmented monitoring procedures.

The reliability of the resulting analysis remained dependent on the quality of the measurements and the validity of the underlying dispersion model. Source localization performed better in open or less obstructed environments, whereas lower accuracy at the wastewater facility reflected the limitations of steady-state Gaussian plume inversion in built environments. Buildings, vegetation, facility structures, unstable winds, and insufficient spatial coverage can distort plume transport and reduce the ability of the model to constrain a unique source. Results produced under such conditions should therefore be interpreted as decision-support information and verified through additional field measurements rather than treated as autonomous final decisions.

Future work will focus on improving the reliability of the framework under more complex monitoring conditions. The dispersion modeling and validity assessment will be further improved to help the system recognize when the available data are insufficient for reliable analysis. Additional self-checking mechanisms may also reduce errors during tool execution and result interpretation.

Another important direction is to systematically evaluate how the choice of LLM affects overall agent performance. Future studies will compare different LLM families and model sizes to separate the contribution of the underlying model from that of the agent scaffolding and integrated tools. Domain-specific post-training may also improve tool usage and reduce the failure cases observed in this study. The framework will also be optimized for smaller and more portable computing platforms. This could support its integration with remotely operated aerial or ground-based sensing systems in hazardous or difficult-to-access environments.
Beyond methane monitoring, the same architecture may be adapted to other gas-sensing applications, including our previously developed CO sensing platform \cite{wengenvironment}, as well as H$_2$S detection and other atmospheric monitoring applications.

\subsection*{Data and Code Availability}

The code and data supporting this study are available from the corresponding author upon reasonable request for research and validation purposes. The source code and selected anonymized test datasets will be made publicly available upon publication. Some facility coordinates and site-identifying information are not publicly available due to site and partner confidentiality requirements.
\begin{acknowledgement}

During the preparation of this manuscript, the author used ChatGPT (GPT-5.5, OpenAI, 2026 version) and Gemini (Google, 2026 version) for language editing, figure generation and refinement, and figure caption editing. The author carefully reviewed, verified, and edited all AI-assisted outputs and takes full responsibility for the content of this publication.

\end{acknowledgement}


\bibliography{Reference}

@article{duncan2015methane,
  title={Does methane pose significant health and public safety hazards?---A review},
  author={Duncan, Ian J.},
  journal={Environmental Geosciences},
  volume={22},
  number={3},
  pages={85--96},
  year={2015},
  doi={10.1306/eg.06191515005}
}

@article{cheng2026leveraging,
  title={Leveraging Large Language Models for Contextual Prioritization of Contaminants of Emerging Concern in Chemical Mixtures},
  author={Cheng, Fei and Li, Qianhui and He, Liwei and Li, Huizhen and Brooks, Bryan W and Yu, Zhiqiang and You, Jing},
  journal={Environmental Science \& Technology},
  volume={60},
  number={15},
  pages={11380--11391},
  year={2026},
  publisher={ACS Publications}
}

@article{nisbet2020methane,
  title={Methane mitigation: methods to reduce emissions, on the path to the Paris agreement},
  author={Nisbet, EG and Fisher, RE and Lowry, D and France, JL and Allen, G and Bakkaloglu, S and Broderick, TJ and Cain, M and Coleman, M and Fernandez, J and others},
  journal={Reviews of Geophysics},
  volume={58},
  number={1},
  pages={e2019RG000675},
  year={2020},
  publisher={Wiley Online Library},
  doi={10.1029/2019RG000675},
  url={https://agupubs.onlinelibrary.wiley.com/doi/full/10.1029/2019RG000675}
}

@article{zhai2026waterrag,
  title={{WaterRAG}: A Multiagent Retrieval-Augmented Generation Framework to Support Water Industry Transitions to Net-Zero},
  author={Zhai, Mudi and Zeng, Qingyun and Qiu, Ruihong and Li, Jiaying and Zhu, Qixiang and Waite, T. David and Ni, Bing-Jie and Duan, Haoran},
  journal={Environmental Science \& Technology},
  volume={60},
  number={15},
  pages={11529--11541},
  year={2026},
  doi={10.1021/acs.est.5c15806}
}

@inproceedings{li2026natural,
  title={Natural Language to {DGGS}-Aware Methane Insights with a Multi-{LLM}-Agent Framework},
  author={Li, Mingke Erin and Liang, Steve},
  booktitle={Proceedings of the 9th Conference on Spatial Knowledge and Information (SKI Canada 2026)},
  address={Banff, Alberta, Canada},
  year={2026}
}

@article{saunois2025global,
  title={Global methane budget 2000--2020},
  author={Saunois, Marielle and Martinez, Adrien and Poulter, Benjamin and Zhang, Zhen and Raymond, Peter A and Regnier, Pierre and Canadell, Josep G and Jackson, Robert B and Patra, Prabir K and Bousquet, Philippe and others},
  journal={Earth System Science Data},
  volume={17},
  number={5},
  pages={1873--1958},
  year={2025},
  publisher={Copernicus GmbH}
}

@article{zhou2021mobile,
  title={Mobile measurement system for the rapid and cost-effective surveillance of methane and volatile organic compound emissions from oil and gas production sites},
  author={Zhou, Xiaochi and Peng, Xiao and Montazeri, Amir and McHale, Laura E and Ga{\ss}ner, Simon and Lyon, David R and Yalin, Azer P and Albertson, John D},
  journal={Environmental science \& technology},
  volume={55},
  number={1},
  pages={581--592},
  year={2021},
  publisher={ACS Publications}
}

@article{jackson2024human,
  title={Human activities now fuel two-thirds of global methane emissions},
  author={Jackson, RB and Saunois, M and Martinez, A and Canadell, JG and Yu, X and Li, M and Poulter, B and Raymond, PA and Regnier, P and Ciais, P and others},
  journal={Environmental Research Letters},
  volume={19},
  number={10},
  pages={101002},
  year={2024},
  publisher={IOP Publishing}
}

@article{brandt2014methane,
  title={Methane leaks from North American natural gas systems},
  author={Brandt, Adam R and Heath, GA and Kort, EA and O'Sullivan, Francis and P{\'e}tron, Gabrielle and Jordaan, Sarah M and Tans, P and Wilcox, Jennifer and Gopstein, AM and Arent, Doug and others},
  journal={Science},
  volume={343},
  number={6172},
  pages={733--735},
  year={2014},
  publisher={American Association for the Advancement of Science}
}

@article{alvarez2018assessment,
  title={Assessment of methane emissions from the US oil and gas supply chain},
  author={Alvarez, Ram{\'o}n A and Zavala-Araiza, Daniel and Lyon, David R and Allen, David T and Barkley, Zachary R and Brandt, Adam R and Davis, Kenneth J and Herndon, Scott C and Jacob, Daniel J and Karion, Anna and others},
  journal={Science},
  volume={361},
  number={6398},
  pages={186--188},
  year={2018},
  publisher={American Association for the Advancement of Science}
}

@article{weller2018vehicle,
  title={Vehicle-based methane surveys for finding natural gas leaks and estimating their size: Validation and uncertainty},
  author={Weller, Zachary D and Roscioli, Joseph R and Daube, W Conner and Lamb, Brian K and Ferrara, Thomas W and Brewer, Paul E and von Fischer, Joseph C},
  journal={Environmental science \& technology},
  volume={52},
  number={20},
  pages={11922--11930},
  year={2018},
  publisher={ACS Publications}
}

@article{chen2023assessing,
  title={Assessing detection efficiencies for continuous methane emission monitoring systems at oil and gas production sites},
  author={Chen, Qining and Schissel, Colette and Kimura, Yosuke and McGaughey, Gary and McDonald-Buller, Elena and Allen, David T},
  journal={Environmental science \& technology},
  volume={57},
  number={4},
  pages={1788--1796},
  year={2023},
  publisher={ACS Publications}
}

@article{chen2026leveraging,
  title={Leveraging LLMs for Environmental Complexity: Structured Fine-Tuning Data Sets and Deployment Strategies},
  author={Chen, Chuke and Li, Nan and Qi, Jianchuan and Chang, Huimin and Shi, Wenjie and Xie, Jinliang and Yuan, Jiayi and Yang, Hang and Guo, Jing and Xu, Changqing and others},
  journal={Environmental Science \& Technology},
  volume={60},
  number={1},
  pages={497--509},
  year={2026},
  publisher={ACS Publications}
}

@article{erland2022recent,
  title={Recent advances toward transparent methane emissions monitoring: a review},
  author={Erland, Broghan M and Thorpe, Andrew K and Gamon, John A},
  journal={Environmental Science \& Technology},
  volume={56},
  number={23},
  pages={16567--16581},
  year={2022},
  publisher={ACS Publications}
}

@article{schwietzke2017improved,
  title={Improved mechanistic understanding of natural gas methane emissions from spatially resolved aircraft measurements},
  author={Schwietzke, Stefan and P{\'e}tron, Gabrielle and Conley, Stephen and Pickering, Cody and Mielke-Maday, Ingrid and Dlugokencky, Edward J and Tans, Pieter P and Vaughn, Tim and Bell, Clay and Zimmerle, Daniel and others},
  journal={Environmental Science \& Technology},
  volume={51},
  number={12},
  pages={7286--7294},
  year={2017},
  publisher={ACS Publications}
}

@article{albertson2016mobile,
  title={A mobile sensing approach for regional surveillance of fugitive methane emissions in oil and gas production},
  author={Albertson, John D and Harvey, Tierney and Foderaro, Greg and Zhu, Pingping and Zhou, Xiaochi and Ferrari, Silvia and Amin, M Shahrooz and Modrak, Mark and Brantley, Halley and Thoma, Eben D},
  journal={Environmental science \& technology},
  volume={50},
  number={5},
  pages={2487--2497},
  year={2016},
  publisher={ACS Publications}
}

@article{jacob2016satellite,
  title={Satellite observations of atmospheric methane and their value for quantifying methane emissions},
  author={Jacob, Daniel J and Turner, Alexander J and Maasakkers, Joannes D and Sheng, Jianxiong and Sun, Kang and Liu, Xiong and Chance, Kelly and Aben, Ilse and McKeever, Jason and Frankenberg, Christian},
  journal={Atmospheric Chemistry and Physics},
  volume={16},
  number={22},
  pages={14371--14396},
  year={2016},
  publisher={Copernicus Publications G{\"o}ttingen, Germany}
}

@article{sherwin2022single,
  title={Single-blind validation of space-based point-source methane emissions detection and quantification},
  author={Sherwin, Evan David and Rutherford, Jeffrey S and Chen, Yuanlei and Aminfard, Sam and Kort, Eric A and Jackson, Robert B and Brandt, Adam R},
  year={2022},
  publisher={EarthArXiv}
}

@article{thorpe2023attribution,
  title={Attribution of individual methane and carbon dioxide emission sources using EMIT observations from space},
  author={Thorpe, Andrew K and Green, Robert O and Thompson, David R and Brodrick, Philip G and Chapman, John W and Elder, Clayton D and Irakulis-Loitxate, Itziar and Cusworth, Daniel H and Ayasse, Alana K and Duren, Riley M and others},
  journal={Science advances},
  volume={9},
  number={46},
  pages={eadh2391},
  year={2023},
  publisher={American Association for the Advancement of Science}
}

@article{ijzermans2024long,
  title={Long-term continuous monitoring of methane emissions at an oil and gas facility using a multi-open-path laser dispersion spectrometer},
  author={IJzermans, Rutger and Jones, Matthew and Weidmann, Damien and van de Kerkhof, Bas and Randell, David},
  journal={Scientific Reports},
  volume={14},
  number={1},
  pages={623},
  year={2024},
  publisher={Nature Publishing Group UK London}
}

@article{ball2025performance,
  title={Performance Evaluation of Fixed-Point Continuous Monitoring Systems: Influence of Averaging Time in Complex Emission Environments},
  author={Ball, David and Eichenlaub, Nathan and Lashgari, Ali},
  journal={Sensors},
  volume={25},
  number={9},
  pages={2801},
  year={2025},
  publisher={MDPI}
}

@article{metzger2025framework,
  title={A Framework for Optimizing Continuous Methane Monitoring System Configuration for Minimal Blind Time: Application and Insights from over 100 Operational Oil and Gas Facilities},
  author={Metzger, Noah and Lashgari, Ali and Esmail, Umair and Ball, David and Eichenlaub, Nathan},
  journal={ACS ES\&T Air},
  volume={2},
  number={8},
  pages={1439--1453},
  year={2025},
  publisher={ACS Publications}
}

@article{coburn2018regional,
  title={Regional trace-gas source attribution using a field-deployed dual frequency comb spectrometer},
  author={Coburn, Sean and Alden, Caroline B and Wright, Robert and Cossel, Kevin and Baumann, Esther and Truong, Gar-Wing and Giorgetta, Fabrizio and Sweeney, Colm and Newbury, Nathan R and Prasad, Kuldeep and others},
  journal={Optica},
  volume={5},
  number={4},
  pages={320--327},
  year={2018},
  publisher={Optical Society of America}
}

@article{yang2025assessing,
  title={Assessing the performance of point sensor continuous monitoring systems at midstream natural gas compressor stations},
  author={Yang, Shuting Lydia and Ravikumar, Arvind P},
  journal={ACS ES\&T Air},
  volume={2},
  number={4},
  pages={466--475},
  year={2025},
  publisher={ACS Publications}
}

@article{yan2025machine,
  title={Machine Learning-Enhanced NDIR Methane Sensing Solution for Robust Outdoor Continuous Monitoring Applications},
  author={Yan, Yang and Mijiddorj, Lkhanaajav and Beringer, Tyler and Mijiddorj, Bilguunzaya and Ho, Alex and Weng, Binbin},
  journal={Sensors},
  volume={25},
  number={24},
  pages={7691},
  year={2025},
  publisher={MDPI}
}

@article{bell2023performance,
  title={Performance of continuous emission monitoring solutions under a single-blind controlled testing protocol},
  author={Bell, Clay and Ilonze, Chiemezie and Duggan, Aidan and Zimmerle, Daniel},
  journal={Environmental science \& technology},
  volume={57},
  number={14},
  pages={5794},
  year={2023}
}

@article{zhou2026aimnet,
  title={AIMNET: An IoT-Empowered Digital Twin for Continuous Gas Emission Monitoring and Early Hazard Detection},
  author={Zhou, Zifan and Wang, Xuan and Yan, Yang and Mijiddorj, Lkhanaajav and Ding, Yu and Beringer, Tyler and Khiabani, Parisa Masnadi and Jentner, Wolfgang G and Hu, Xiao-Ming and Wang, Chenghao and others},
  journal={IEEE Internet of Things Magazine},
  year={2026},
  publisher={IEEE}
}

@mastersthesis{yan2025development,
  title={Development of a Machine-Learning Enhanced High Performance Methane Sensing Instrument for Field Applications},
  author={Yan, Yang},
  year={2025},
  school={University of Oklahoma--Graduate College}
}

@inproceedings{hu2026multi,
  title={Multi-Sensor and Multi-Model Investigation of Methane Plumes from a Wastewater Treatment Plant to Improve Emission Inversion during the Morning Boundary Layer Transition},
  author={Hu, Xiao-Ming and Ding, Yu and Yan, Yang and Wang, Chenghao and Weng, Binbin and Hardeman, Steve and Xue, Ming},
  booktitle={106th AMS Annual Meeting},
  year={2026},
  organization={AMS}
}

@article{weng2024road,
  title={The road to climate change mitigation via methane emissions monitoring},
  author={Weng, Binbin},
  journal={Nature Reviews Electrical Engineering},
  volume={1},
  number={2},
  pages={69--70},
  year={2024},
  publisher={Nature Publishing Group UK London}
}

@article{achiam2023gpt,
  title={Gpt-4 technical report},
  author={Achiam, Josh and Adler, Steven and Agarwal, Sandhini and Ahmad, Lama and Akkaya, Ilge and Aleman, Florencia Leoni and Almeida, Diogo and Altenschmidt, Janko and Altman, Sam and Anadkat, Shyamal and others},
  journal={arXiv preprint arXiv:2303.08774},
  year={2023}
}

@inproceedings{wu2024autogen,
  title={Autogen: Enabling next-gen LLM applications via multi-agent conversations},
  author={Wu, Qingyun and Bansal, Gagan and Zhang, Jieyu and Wu, Yiran and Li, Beibin and Zhu, Erkang and Jiang, Li and Zhang, Xiaoyun and Zhang, Shaokun and Liu, Jiale and others},
  booktitle={First conference on language modeling},
  year={2024}
}

@article{guo2024large,
  title={Large language model based multi-agents: A survey of progress and challenges},
  author={Guo, Taicheng and Chen, Xiuying and Wang, Yaqi and Chang, Ruidi and Pei, Shichao and Chawla, Nitesh V and Wiest, Olaf and Zhang, Xiangliang},
  journal={arXiv preprint arXiv:2402.01680},
  year={2024}
}

@article{wang2022multiscale,
  title={Multiscale methane measurements at oil and gas facilities reveal necessary frameworks for improved emissions accounting},
  author={Wang, Jiayang Lyra and Daniels, William S and Hammerling, Dorit M and Harrison, Matthew and Burmaster, Kaylyn and George, Fiji C and Ravikumar, Arvind P},
  journal={Environmental science \& technology},
  volume={56},
  number={20},
  pages={14743--14752},
  year={2022},
  publisher={ACS Publications}
}

@article{duan2024exploration,
  title={Exploration of llm multi-agent application implementation based on langgraph+ crewai},
  author={Duan, Zhihua and Wang, Jialin},
  journal={arXiv preprint arXiv:2411.18241},
  year={2024}
}

@article{han2024llm,
  title={LLM multi-agent systems: Challenges and open problems},
  author={Han, Shanshan and Zhang, Qifan and Jin, Weizhao and Xu, Zhaozhuo},
  journal={arXiv preprint arXiv:2402.03578},
  year={2024}
}

@article{vaswani2017attention,
  title={Attention is all you need},
  author={Vaswani, Ashish and Shazeer, Noam and Parmar, Niki and Uszkoreit, Jakob and Jones, Llion and Gomez, Aidan N and Kaiser, {\L}ukasz and Polosukhin, Illia},
  journal={Advances in neural information processing systems},
  volume={30},
  year={2017}
}

@article{brown2020language,
  title={Language models are few-shot learners},
  author={Brown, Tom and Mann, Benjamin and Ryder, Nick and Subbiah, Melanie and Kaplan, Jared D and Dhariwal, Prafulla and Neelakantan, Arvind and Shyam, Pranav and Sastry, Girish and Askell, Amanda and others},
  journal={Advances in neural information processing systems},
  volume={33},
  pages={1877--1901},
  year={2020}
}

@article{minaee2024large,
  title={Large language models: A survey},
  author={Minaee, Shervin and Mikolov, Tomas and Nikzad, Narjes and Chenaghlu, Meysam and Socher, Richard and Amatriain, Xavier and Gao, Jianfeng},
  journal={arXiv preprint arXiv:2402.06196},
  year={2024}
}

@article{yuvaraj2026high,
  title={High-resolution modeling of methane plumes: Validation and sensitivity experiments to explore emission quantification approaches},
  author={Yuvaraj, Rakesh and Lauvaux, Thomas and Abdallah, Charbel and Ciais, Philippe and Akani Guery, Julian and Bonne, Jean-Louis and Groshenry, Alexis and Hoang, Ngoc Minh and Joly, Lilian},
  journal={Environmental Science \& Technology},
  volume={60},
  number={5},
  pages={4029--4041},
  year={2026},
  publisher={ACS Publications}
}

@article{li2023camel,
  title={Camel: Communicative agents for" mind" exploration of large language model society},
  author={Li, Guohao and Hammoud, Hasan and Itani, Hani and Khizbullin, Dmitrii and Ghanem, Bernard},
  journal={Advances in neural information processing systems},
  volume={36},
  pages={51991--52008},
  year={2023}
}

@inproceedings{hong2023metagpt,
  title={MetaGPT: Meta programming for a multi-agent collaborative framework},
  author={Hong, Sirui and Zhuge, Mingchen and Chen, Jonathan and Zheng, Xiawu and Cheng, Yuheng and Wang, Jinlin and Zhang, Ceyao and Wang, Zili and Yau, Steven Ka Shing and Lin, Zijuan and others},
  booktitle={The twelfth international conference on learning representations},
  year={2023}
}

@inproceedings{huang2025middle,
  title={A Middle Path for On-Premises LLM Deployment: Preserving Privacy Without Sacrificing Model Confidentiality},
  author={Huang, Hanbo and Li, Yihan and Jiang, Bowen and Jiang, Bo and Liu, Lin and Liu, Zhuotao and Sun, Ruoyu and Liang, Shiyu},
  booktitle={Proceedings of the 2025 Conference on Empirical Methods in Natural Language Processing},
  pages={8332--8370},
  year={2025}
}

@article{grattafiori2024llama,
  title={The llama 3 herd of models},
  author={Grattafiori, Aaron and Dubey, Abhimanyu and Jauhri, Abhinav and Pandey, Abhinav and Kadian, Abhishek and Al-Dahle, Ahmad and Letman, Aiesha and Mathur, Akhil and Schelten, Alan and Vaughan, Alex and others},
  journal={arXiv preprint arXiv:2407.21783},
  year={2024}
}

@article{yang2025qwen3,
  title={Qwen3 technical report},
  author={Yang, An and Li, Anfeng and Yang, Baosong and Zhang, Beichen and Hui, Binyuan and Zheng, Bo and Yu, Bowen and Gao, Chang and Huang, Chengen and Lv, Chenxu and others},
  journal={arXiv preprint arXiv:2505.09388},
  year={2025}
}

@misc{ollamaGithub2026,
  title={Ollama: Get up and running with large language models},
  author={{Ollama}},
  howpublished={\url{https://github.com/ollama/ollama}},
  year={2026}
}

@article{wengenvironment,
  title={An Environment-Adaptive Low-Power IoT Architecture for Portable Smoking Detection with Real-Time Data Quality Assurance},
  author={Weng, Binbin and Mijiddorj, Bilguunzaya and Beringer, Tyler and Mijiddorj, Lkhanaajav and Yang, Yan and Ho, Alex and Hassan, Erum}
}

@article{bai2025qwen3vl,
  title={Qwen3-VL Technical Report},
  author={Bai, Shuai and Cai, Yuxuan and Chen, Ruizhe and Chen, Keqin and Chen, Xionghui and Cheng, Zesen and Deng, Lianghao and Ding, Wei and Gao, Chang and Ge, Chunjiang and others},
  journal={arXiv preprint arXiv:2511.21631},
  year={2025}
}

\newpage

\begin{center}
    {\LARGE\bfseries Supporting Information}\\[1.2em]

{\large\bfseries
A Locally Deployable Tool-Grounded LLM Multi-agent Framework for Automating Methane Emission Analysis and Reporting
}\\[1em]

    Yang Yan,$^{\dagger}$ 
Zifan Zhou,$^{\ddagger}$ 
Xuan Wang,$^{\ddagger}$ 
Erum Hassan,$^{\dagger}$ 
Bilguunzaya Mijiddorj,$^{\dagger}$

\vspace{0.5em}

Jie Cao,$^{\P}$ 
Bin Li,$^{\ddagger}$ 
and Binbin Weng$^{*,\dagger}$

\vspace{1em}

{\itshape
$^{\dagger}$School of Electrical and Computer Engineering, University of Oklahoma, Norman, OK,\\
73019 USA
}

\vspace{0.5em}

{\itshape
$^{\ddagger}$Department of Electrical Engineering, Pennsylvania State University, University Park,\\
PA, 16801 USA
}

\vspace{0.5em}

{\itshape
$^{\P}$School of Computer Science, University of Oklahoma, Norman, OK, 73019 USA
}

\vspace{1em}

E-mail: binbinweng@ou.edu
\end{center}

\vspace{1em}
\setcounter{table}{0}
\renewcommand{\thetable}{S\arabic{table}}
\renewcommand{\theHtable}{S\arabic{table}}

\section*{S1.Scoring Rubric}
The LLM-agent outputs were evaluated using a 0--10 rubric. The score was assigned based on task completion, scientific validity, data handling, visualization and reporting quality, and decision-support usefulness.

\begin{table}[htbp]
\centering
\caption{General scoring rubric for evaluating LLM-agent outputs.}
\label{tab:S_score_rubric}
\begin{tabular}{p{0.18\linewidth}p{0.74\linewidth}}
\hline
Score range & Description \\
\hline
9--10 & Excellent: accurately presents methane data, meteorological inputs, source location, model-derived emission rate, plume results, uncertainty, and validity information. Figures and conclusions are consistent and require only minor edits. \\

7--8 & Good: main analytical results are correct and usable, with only minor omissions or unclear details in weather sources, uncertainty, validity, figures, or recommendations. \\

5--6 & Fair: major outputs are identifiable, but important information is incomplete, inconsistent, or requires expert revision before use. \\

3--4 & Poor: major omissions or errors in methane units, meteorological inputs, source localization, plume results, figures, or conclusions substantially limit usability. \\

1--2 & Very poor: mostly incorrect or incomplete, with unsupported conclusions, physically inconsistent results, or unreadable report content. \\

0 & Failed: no valid report, failed execution, unreadable output, or content unrelated to the methane-monitoring task. \\
\hline
\end{tabular}
\end{table}

\section*{S2. Example Prompts}

Fifty representative prompts were used to evaluate task interpretation, parameter extraction, workflow selection, and tool coordination. The prompts covered field planning, data processing, source and plume analysis, emission estimation, and report generation. Actual coordinates were replaced with “[latitude, longitude]” to protect facility locations.
\begin{table}[H]
\centering
\caption{Example prompts for evaluating methane field-test route planning}
\label{tab:S_prompts_routing}
\begin{tabular}{p{0.08\linewidth}p{0.84\linewidth}}
\hline
No. & Example prompt \\
\hline
1 & I want to conduct a methane field test around [latitude, longitude] tomorrow morning. What is the best time to start? \\

2 & I want to conduct a methane field test around [latitude, longitude] this week. Which day and time would be most suitable? \\

3 & I plan to test methane emissions around [latitude, longitude] on July 5. What would be the best time of day? \\

4 & I want to conduct a field test around [latitude, longitude] this afternoon. Are the expected weather and wind conditions suitable? \\

5 & I want to conduct a methane survey around [latitude, longitude] next week. Recommend the best date and time based on the weather forecast. \\

6 & I want to conduct a methane field test around [latitude, longitude]. What route should I follow? \\

7 & I want to test methane emissions around the landfill sometime this week. Which monitoring route would be most suitable based on the expected wind direction? \\

8 & I want to conduct a methane field test around the oil and gas facility on July 5. Which route should I take? \\

9 & I plan to conduct a mobile methane survey around [latitude, longitude] tomorrow morning. Generate a route that covers the most likely downwind detection areas. \\

10 & I want to conduct a methane field test around [latitude, longitude] sometime this week. Recommend the best day and time and generate an appropriate monitoring route. \\
\hline
\end{tabular}
\end{table}
\begin{table}[H]
\centering
\caption{Example user prompts for methane-data retrieval, preprocessing, and input validation}
\label{tab:S_prompts_planning}
\begin{tabular}{p{0.08\linewidth}p{0.84\linewidth}}
\hline
No. & Example prompt \\
\hline
11 & Plot CH$_4$, humidity, and temperature from the uploaded CSV file on the same time axis. \\

12 & Plot CH$_4$, CO$_2$, and H$_2$O measurements from the uploaded CSV file for comparison. \\

13 & Filter out abnormal methane measurements from the uploaded dataset and plot the cleaned data. \\

14 & Process the uploaded raw AIMNet sensor data and plot the methane measurements before and after preprocessing. \\

15 & Collect ten minutes of data from AIMNet devices 001, 002, and 003 and plot their methane measurements on the same figure. \\

16 & Compare the methane measurements from AIMNet devices 001, 002, and 003. Identify which device reports the highest values and whether any device shows abnormal measurements. \\

17 & The LI-COR LI-7810 measurements are 56 seconds behind the AIMNet measurements. Correct the time offset and plot the aligned data. \\

18 & The LI-COR LI-7810 measurements are 56 seconds behind, and the LI-COR LI-7700 measurements are 21 seconds behind. Correct both time offsets and plot the aligned LI-7810, LI-7700, and AIMNet device measurements on the same figure. \\

19 & Check the uploaded AIMNet dataset for missing values, duplicate timestamps, invalid GPS records, and abnormal methane measurements. \\

20 & Compare the processed AIMNet methane measurements with the corresponding LI-COR measurements and summarize the differences among the sensors. \\
\hline
\end{tabular}
\end{table}

\begin{table}[H]
\centering
\caption{Example prompts for known-source plume reconstruction and emission-rate estimation.}
\label{tab:S_prompts_data_processing}
\begin{tabular}{p{0.08\linewidth}p{0.84\linewidth}}
\hline
No. & Example prompt \\
\hline
21 & The methane source is located at [latitude, longitude]. The wind speed was 6 mph and the wind direction was 150 degrees. Methane enhancement was 1 ppm at 5 m and 200 ppb at 10 m. Reconstruct the plume. \\

22 & I tested a methane source at [latitude, longitude]. The wind speed was 3 m/s and the wind direction was 220 degrees. Methane enhancement was 1.2 ppm at 5 m, 500 ppb at 10 m, and 150 ppb at 20 m. Estimate the emission rate and plot the plume. \\

23 & I tested a methane source at [latitude, longitude] on July 5, 2026, at 10:00 a.m. Methane enhancement was 800 ppb at 8 m and 300 ppb at 15 m. Retrieve the weather conditions and reconstruct the plume. \\

24 & I tested a methane source at [latitude, longitude] on January 20, 2026, at 2:00 p.m. Methane concentrations were 3.0 ppm at 5 m and 2.3 ppm at 10 m, with a background of 2.0 ppm. Retrieve the weather conditions and plot the plume. \\

25 & The methane source is located at [latitude, longitude]. The wind speed was 3.1 m/s and the wind direction was 295 degrees. Methane enhancement was 2400 ppb at 6 m and 700 ppb at 12 m. Reconstruct the plume. \\

26 & I tested a methane source at [latitude, longitude]. The wind speed was 2.5 m/s and the wind direction was 45 degrees. Methane concentrations were 3.4 ppm at 4 m and 2.5 ppm at 9 m, with a background of 2.1 ppm. Estimate the emission rate. \\

27 & I tested a methane source at [latitude, longitude] on August 3, 2026, at 9:00 a.m. Methane enhancement was 1.5 ppm at 5 m, 600 ppb at 10 m, and 200 ppb at 20 m. Retrieve the weather conditions and reconstruct the plume. \\

28 & I uploaded methane and GPS measurements collected around a known source at [latitude, longitude]. Use the measurement timestamps to retrieve the weather conditions and generate a plume map. \\

29 & The source is located at [latitude, longitude]. The wind speed was 4 m/s and the wind direction was 180 degrees. Methane enhancement was 900 ppb at 5 m, 450 ppb at 10 m, and 120 ppb at 20 m. Plot the plume and estimate the emission rate. \\

30 & I tested a methane source at [latitude, longitude] on June 15, 2026, at 11:00 a.m. Methane concentrations were 2.9 ppm at 5 m, 2.4 ppm at 10 m, and 2.1 ppm at 20 m, with a background of 2.0 ppm. Retrieve the weather conditions and reconstruct the plume. \\
\hline
\end{tabular}
\end{table}

\begin{table}[H]
\centering
\caption{Example user prompts for inverse methane-source localization and emission analysis}
\label{tab:S_prompts_plume_analysis}
\begin{tabular}{p{0.08\linewidth}p{0.84\linewidth}}
\hline
No. & Example prompt \\
\hline
31 & I conducted a vehicle-based methane survey. Plot the methane measurements along the survey route, identify the suspected source location, and generate a plume map. \\

32 & I uploaded a manual-survey CSV file containing methane concentrations and GPS coordinates. Estimate the most likely source location and reconstruct the plume. \\

33 & I conducted a vehicle survey around [latitude, longitude] on July 5, 2026, at 10:00 a.m. Retrieve the corresponding weather conditions and identify the most likely methane source. \\

34 & I uploaded vehicle-survey data with a wind speed of 3 m/s and a wind direction of 220 degrees. Estimate the source location, emission rate, and plume distribution. \\

35 & Compare the possible source locations around the survey route and identify the location that best matches the observed methane measurements. \\

36 & I uploaded a vehicle-survey CSV file containing methane concentrations, GPS coordinates, and timestamps. Identify the suspected source location, estimate the emission rate, and generate a plume map. \\

37 & Estimate the most likely methane source coordinate and show the model-based source-likelihood region around the predicted location. \\

38 & Determine whether the uploaded methane survey represents a single source or multiple sources. If multiple sources are detected, estimate each source location separately. \\

39 & The uploaded survey contains two spatially separated methane-enhancement regions. Localize the suspected source associated with each region and generate separate plume maps. \\

40 & Analyze the uploaded vehicle-survey data, estimate the source location and emission rate, assess the validity of the localization result, and recommend additional measurements if needed. \\
\hline
\end{tabular}
\end{table}
\begin{table}[H]
\centering
\caption{Example user prompts for validity assessment, report generation, and report revision}
\label{tab:S_prompts_report}
\begin{tabular}{p{0.08\linewidth}p{0.84\linewidth}}
\hline
No. & Example prompt \\
\hline
41 & Generate a methane-monitoring report for the monitoring point at [latitude, longitude]. \\

42 & Generate a report summarizing the results from the last three monitoring points. \\

43 & Generate a report for the field-test route planned on July 10. \\

44 & Prepare a report for the most recent vehicle-based methane survey, including the concentration map, predicted leak location, plume visualization, and emission-rate estimate. \\

45 & Generate a short field report for the manual methane test conducted at [latitude, longitude]. \\

46 & Combine the last three source-localization results into one report and compare their predicted source locations and validity assessments. \\

47 & Add the route plan generated on July 10 to the corresponding methane-monitoring report. \\

48 & Increase the font size of the text, figure labels, and table content in the previous report to improve readability. \\

49 & Change the title of the previous report to ``Methane Monitoring and Emission Analysis Report'' while keeping the remaining content unchanged. \\

50 & Regenerate the previous report with a shorter conclusion and clear separation between measured results and model-derived predictions. \\
\hline
\end{tabular}
\end{table}

\section{S2. Example of a Successful Methane Monitoring Report}
This figure shows an additional successful example of automated report generation. The framework extracted the location information and processed the uploaded CSV measurements. It generated the methane concentration plot and identified the major concentration peaks above the defined threshold. The framework also used multiple measurement points and wind information to generate a Gaussian plume map. The estimated methane release rate was 8.96 g/h. These results were then summarized in the final report with the corresponding figures and interpretations. This example demonstrates the successful execution of a multi-step methane monitoring task.

\begin{figure}[htbp]
    \centering
    \includegraphics[width=1\linewidth]{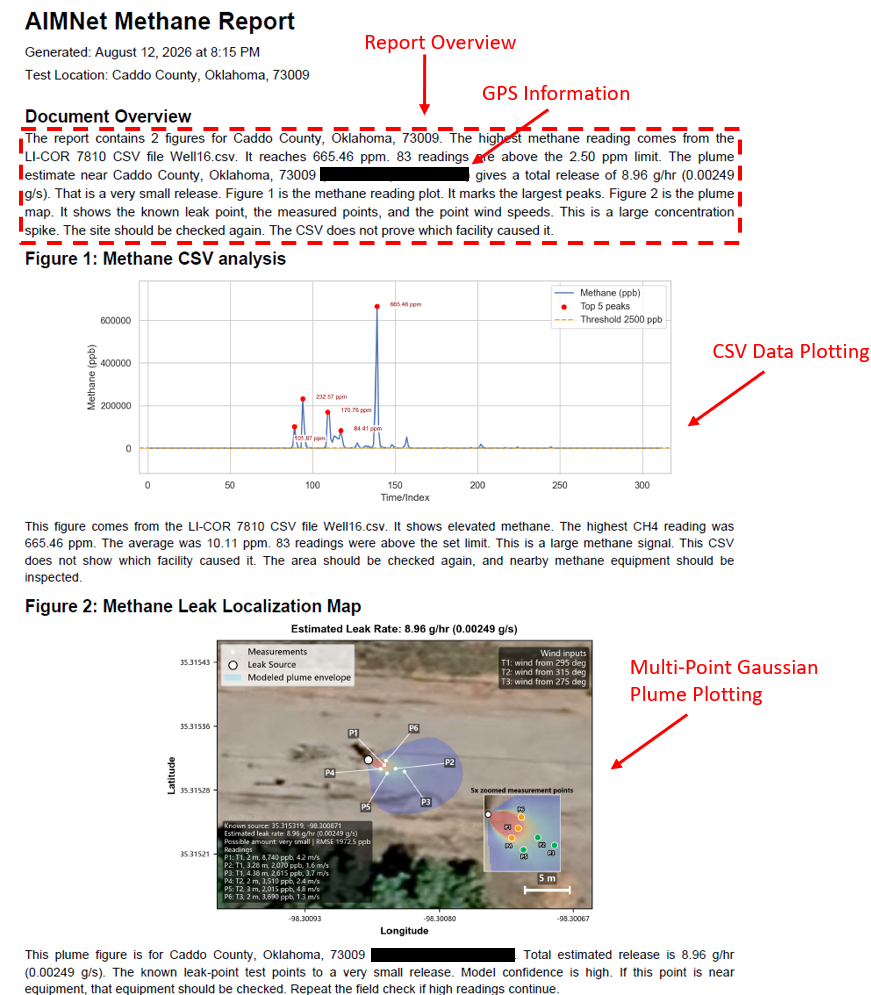}
    \caption{Successful Example of Automated Methane Report Generation.}
    \label{fig:successcase}
\end{figure}

\newpage 
\section*{S4. Failure Cases}

Representative failure cases were included to illustrate the main limitations observed during evaluation. These failures primarily involved misinterpretation of the user’s request, routing to an inappropriate downstream workflow, preprocessing errors such as incorrect methane-column identification, and plume-analysis errors when the retrieved wind speed and direction did not represent the transient local conditions during field measurement.Figures 11–13 present representative examples of these failure modes.

\begin{figure}[htbp]
\centering
\includegraphics[width=0.90\linewidth]{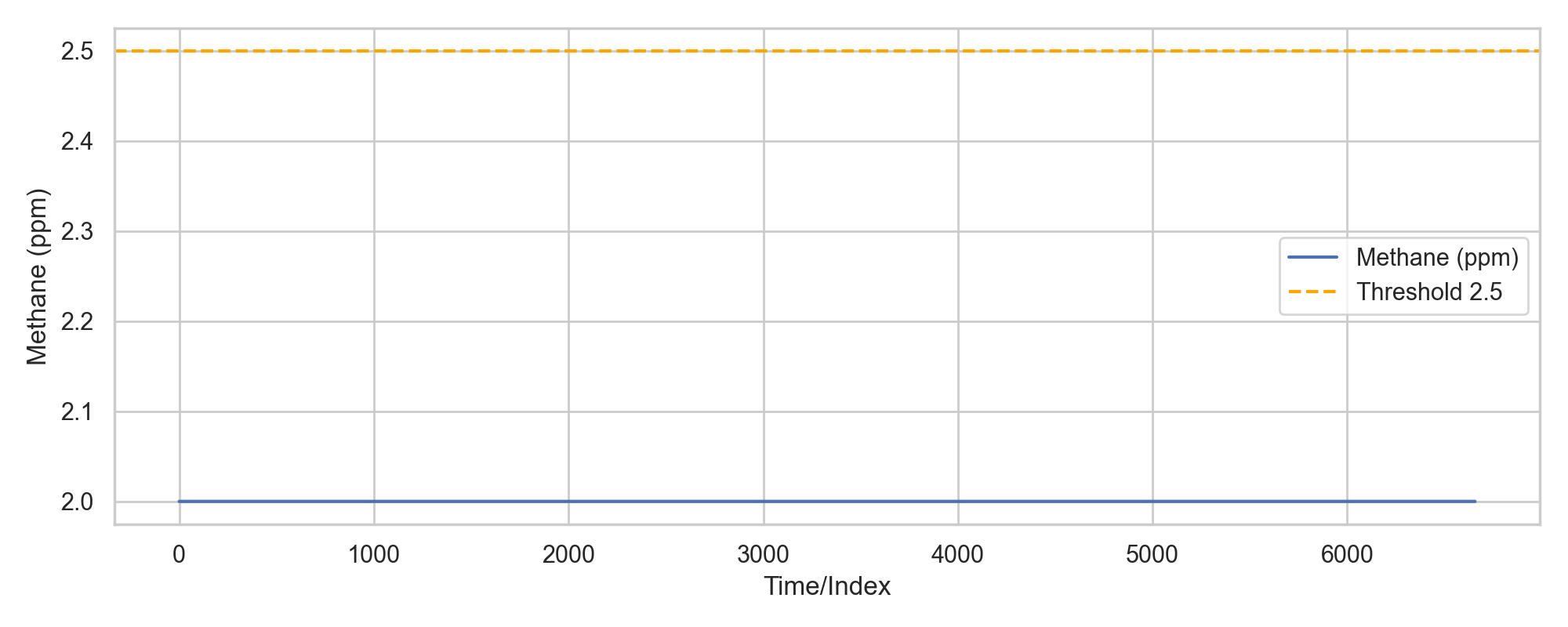}
\caption{Representative failure in automatic methane-column identification and visualization. Because the CSV file contained ambiguous column headers and multiple candidate columns with similar numerical distributions, the preprocessing routine selected the wrong methane field and generated an incorrect methane concentration plot.}
\label{fig:S_failure_case_1}
\end{figure}
\begin{figure}[htbp]
\centering
\includegraphics[width=0.90\linewidth]{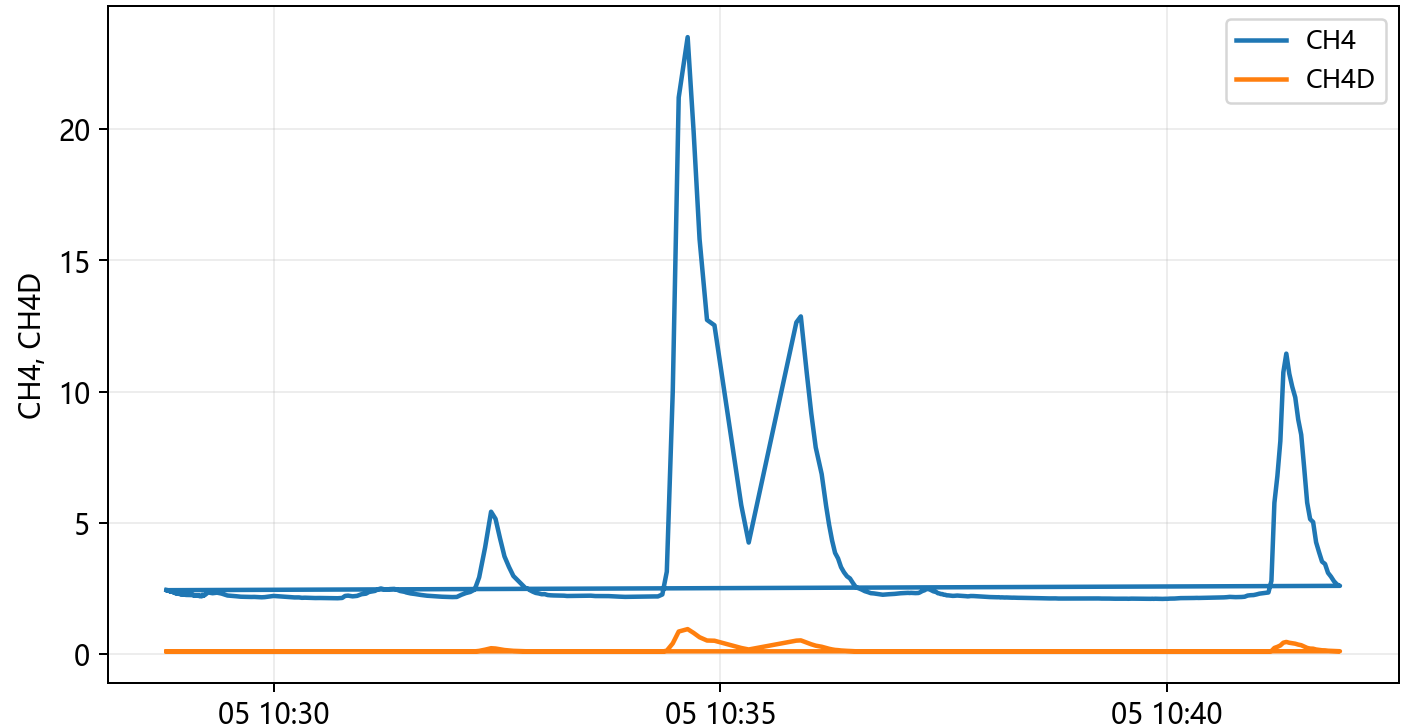}
\caption{Representative failure caused by incomplete time-series data. Missing timestamps and data gaps in the original files caused the agent to incorrectly connect discontinuous measurements during visualization, resulting in a misleading methane concentration plot.}
\label{fig:S_failure_case_3}
\end{figure}

\begin{figure}[htbp]
\centering
\includegraphics[width=0.90\linewidth]{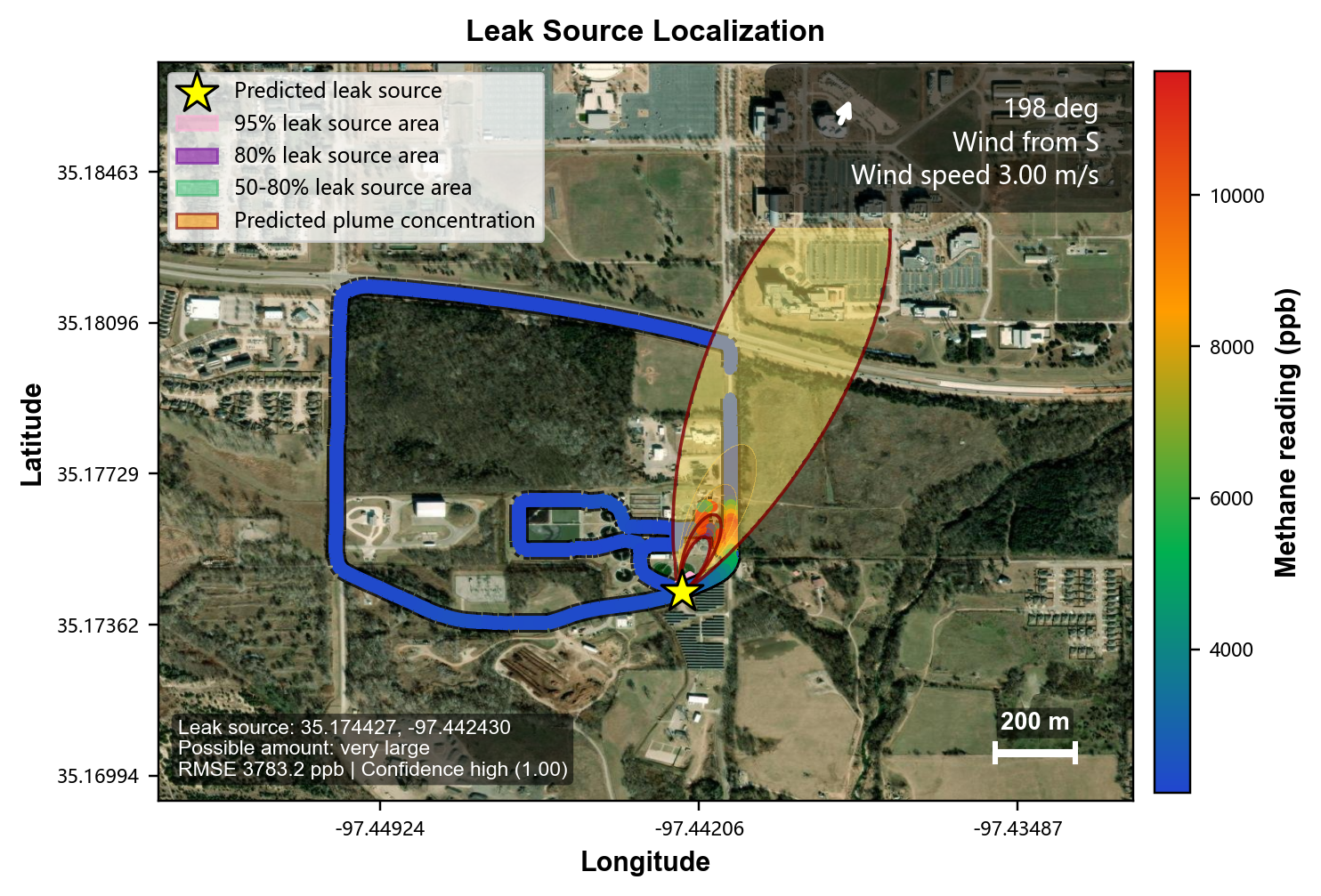}
\caption{Representative plume-reconstruction failure caused by nonrepresentative meteorological inputs. The wind speed and direction retrieved by the Weather Agent did not capture the transient local wind conditions during field measurement, resulting in an inaccurate Gaussian plume visualization and unreliable source-localization and emission-rate estimates.}
\label{fig:S_failure_case_2}
\end{figure}


\end{document}